\documentclass[article,11pt]{article}

\pdfoutput=1

\usepackage{jheppub}
\usepackage{caption}
\usepackage{subcaption}
\usepackage{amssymb}
\usepackage{amsbsy}
\usepackage{amsfonts}
\usepackage{amsmath}
\usepackage{epsfig}
\usepackage{cancel}
\usepackage{slashed}
\usepackage{array}
\usepackage{mathtools}
\usepackage{bm}
\usepackage{braket}
\usepackage{xcolor}
\usepackage{hyperref}

\title{Contributions to $\bm{b \rightarrow s \ell \ell}$ Anomalies from $\bm{R}$-Parity Violating Interactions}

\author[a]{Kevin Earl}
\author[a]{Thomas~Gr\'egoire}

\emailAdd{KevinEarl@cmail.carleton.ca}
\emailAdd{gregoire@physics.carleton.ca}

\affiliation[a]{Ottawa-Carleton Institute for Physics, Department of Physics, Carleton University 1125 Colonel By Drive, Ottawa, K1S 5B6 Canada}
\abstract{
We examine the parameter space of supersymmetric models with $R$-parity violating interactions of the form $\lambda' L Q D^c$ to explain the various anomalies observed in $b \rightarrow s \ell \ell$ transitions. To generate the appropriate operator in the low energy theory, we are led to a region of parameter space where loop contributions dominate. In particular, we concentrate on parameters for which diagrams involving winos, which have not been previously considered, give large contributions. Many different potentially constraining processes are analyzed, including $\tau \rightarrow \mu \mu \mu$, $B_s-\bar{B}_s$ mixing, $B \rightarrow K^{(*)} \nu \bar{\nu}$, $Z$ decays to charged leptons, and direct LHC searches. We find that it is possible to explain the anomalies, but it requires large values of $\lambda'$, which lead to relatively low Landau poles.
} 

\begin{document}

\maketitle
\section{Introduction}
For a number of years, various experiments have reported anomalies in measurements of semileptonic $B$ decays. For example, consider $R_K$ and $R_{K^*}$
\begin{align}
R_{K^{(*)}} = \frac{\text{Br}(B \rightarrow K^{(*)} \mu^+ \mu^-)}{\text{Br}(B \rightarrow K^{(*)} e^+ e^-)}.
\end{align}
As these observables are ratios of branching ratios, they are virtually free of hadronic uncertainties, and thus are excellent tests of lepton flavour universality. The $R_K$ ratio, for the dilepton invariant mass squared range $1$ to $6 \ \text{GeV}^2$, has been measured to be \cite{Aaij:2014ora}
\begin{align}
R_{K} = 0.745^{+0.090}_{-0.074}\text{(stat)}\pm0.036 \text{(syst)}, \ 1 < m_{\ell \ell}^2 < 6 \text{ GeV}^2
\end{align}
by the LHCb collaboration. This represents a $2.6\sigma$ deviation away from the Standard Model prediction, which is 1 with an uncertainty of $\sim 10^{-2}$ \cite{Bordone:2016gaq, Bobeth:2007dw}. Further, the ratio $R_{K^{*}}$ has been measured for two invariant mass squared bins \cite{Aaij:2017vbb}
\begin{align}
R_{K^{*}} = \biggl\{ \begin{matrix*}[l] 0.66^{+0.11}_{-0.07}\text{(stat)} \pm 0.03 \text{(syst)}, & 0.045 < m_{\ell \ell}^2 < 1.1 \text{ GeV}^2 \\
0.69^{+0.11}_{-0.07}\text{(stat)} \pm 0.05 \text{(syst)}, & 1.1 \ \ \ < m_{\ell \ell}^2 < 6.0 \text{ GeV}^2 \\
\end{matrix*}
\end{align}
also by the LHCb collaboration. The Standard Model prediction for these observables varies between $0.878$ and $0.944$ for the low invariant mass squared bin and $0.990$ and $1.010$ for the high invariant mass squared bin \cite{Aaij:2017vbb}. The measured values then represent $2.3\sigma$ and $2.5 \sigma$ deviations for the low and high invariant mass squared bins, respectively. Moreover, some angular distributions also show tension with the Standard Model predictions. In particular, the $P_5'$ observable \cite{Matias:2012xw,Descotes-Genon:2013vna,DescotesGenon:2012zf} in the $B \rightarrow K^* \mu \mu$ decay as measured by Belle \cite{Wehle:2016yoi,Abdesselam:2016llu} and LHCb \cite{Aaij:2015oid,Aaij:2013qta} shows a $2.9 \sigma$ discrepancy \cite{Capdevila:2017bsm}. Finally, LHCb has also observed a deficit exceeding $3\sigma$ in another $b \rightarrow s \mu \mu$  transition, namely the $B_s \rightarrow \phi \mu^+ \mu^-$ decay \cite{Aaij:2015esa,Aaij:2013aln}.

Taken independently, none of these measurements are in dramatic tension with the Standard Model. However, an interesting feature of these anomalies is that model independent analyses \cite{Descotes-Genon:2013wba,Hiller:2014ula,Altmannshofer:2014rta,Descotes-Genon:2015uva,Altmannshofer:2017yso,Capdevila:2017bsm, Bardhan:2017xcc,Ghosh:2017ber} have shown that new physics contributions to effective four-fermi operators can consistently explain nearly all of them. In fact, a fit of the $b \rightarrow s \ell \ell$ transition data to a set of higher dimensional operators shows that new physics is preferred over the Standard Model at the $5\sigma$ level \cite{Capdevila:2017bsm}. Furthermore, these fits unequivocally demonstrate that one potential way to explain these anomalies is to generate new physics contributions to the operator
\begin{align}
(\bar{s} \gamma_\alpha P_L b)(\bar{\mu} \gamma^{\alpha} P_L \mu).
\end{align}

We also note that there are signs of lepton flavour universality violation in the $b \rightarrow c \ell \nu$ transitions as well. Namely, the ratios of branching ratios $R_D$ and $R_{D^*}$
\begin{align} \label{eq:R_D}
R_{D^{(*)}} = \frac{\text{Br}(B \rightarrow D^{(*)}\tau \nu)}{\text{Br}(B \rightarrow D^{(*)}\ell \nu)}
\end{align}
where $\ell = e$ or $\mu$, have been measured by Babar \cite{Lees:2012xj, Lees:2013uzd}, Belle \cite{Huschle:2015rga, Sato:2016svk, Abdesselam:2016xqt, Hirose:2016wfn, Hirose:2017dxl}, and LHCb \cite{Aaij:2015yra} and the results seem to be in tension with the Standard Model \cite{Amhis:2016xyh}. However, in this work we do not focus on these discrepancies, although we do briefly discuss them near the end of the paper.

Many different models featuring new particles, for example leptoquarks (either scalar or vector) that couple to a quark and a lepton, have been proposed to potentially explain these anomalies. Depending on the flavour structure of their couplings, such particles can contribute to the $B$ to $K$ processes, $B$ to $D$ processes, or both \cite{Sakaki:2013bfa, Hiller:2014yaa, Freytsis:2015qca, Bauer:2015knc, Allanach:2015gkd, Dorsner:2016wpm, Das:2016vkr, Becirevic:2016oho, Becirevic:2016yqi, Hiller:2016kry, Becirevic:2017jtw, Alok:2017jaf, Alok:2017sui, Aloni:2017ixa, Assad:2017iib, Calibbi:2017qbu}. In supersymmetric models featuring the $R$-parity violating (RPV) term $\lambda' L Q D^c$ in the superpotential, the squarks are in fact leptoquarks. Therefore, such models provide a natural framework to address the anomalies \cite{Biswas:2014gga, Das:2017kfo, Deshpande:2012rr, Deshpand:2016cpw, Altmannshofer:2017poe}. To explain the anomalies in the $b\rightarrow s \mu\mu$ transition we are led to consider loop level contributions as tree level exchange of squarks lead to four-fermi operators with incorrect chirality structures. In these models, there are various kinds of box diagrams that contribute. One class of diagrams involve only intermediate squarks and were considered in a previous work on leptoquarks \cite{Bauer:2015knc}. In addition, there are diagrams that also involve sleptons which are specific to supersymmetric models. Those contributions were considered in \cite{Das:2017kfo} which found regions of parameters space that could explain the anomalies and avoid constraints. These regions are characterized by large $\lambda'$ couplings and TeV-scale superpartners. As a part of this work, we reexamine this parameter space and find new constraints. Finally, in supersymmetric RPV models, there are diagrams involving winos. These have not been considered previously in the literature with regards to the anomalies. Therefore, in this paper we focus our attention on regions of parameter space where such diagrams give significant contributions. This leads us to a parameter space where the couplings $\lambda'_{223},\lambda'_{233},\lambda'_{323}$, and $ \lambda'_{333}$ are each large. Additionally, the masses of the left-handed squark doublets need to be of order $1 \ \text{TeV}$, while to avoid various experiment constraints the masses of the right-handed sbottom and the left-handed slepton doublets need to be of order $10 \ \text{TeV}$.

This paper is structured as follows. In section \ref{sec:calculations} we compute the contribution of our model to the relevant four-fermi effective operators. We then discuss the region of parameter space which is the focus of our work. In section \ref{sec:constraints}  we present various constraints on the model. In particular, the processes $\tau \rightarrow \mu \mu \mu$, $B_s-\bar{B}_s$ mixing, $B \rightarrow K^{(*)} \nu \bar{\nu}$, $Z$ decays to charged leptons, direct LHC searches, and the presence of Landau poles are examined. Finally, we present our results in section \ref{sec:results} and we conclude in section \ref{sec:conclusion}.

\section{Setup and calculations}\label{sec:calculations}

The effects of new physics on the decay $b \rightarrow s \ell \ell$ can be encoded in contributions to higher dimensional operators. Specifically, the low energy effective Hamiltonian is often parametrized as
\begin{align}
\mathcal{H}_{\text{eff}} = -\frac{4G_F}{\sqrt{2}}V_{tb}V^*_{ts}\frac{\alpha}{4\pi}\sum_{\ell = e,\mu}(C_9^\ell O_9^\ell + C_{10}^\ell O_{10}^\ell + C_{9}^{\prime \ell} O_{9}^{\prime \ell} + C_{10}^{\prime \ell} O_{10}^{\prime \ell}) + \text{h.c.},
\end{align}
where $G_F$ is Fermi's constant, $V_{ij}$ is the CKM matrix, $\alpha$ is the fine-structure constant, and
\begin{alignat}{2}
O_9^\ell &= (\bar{s}\gamma_\alpha P_L b)(\bar{\ell}\gamma^\alpha \ell), \qquad &&O_{9}^{\prime \ell} = (\bar{s}\gamma_\alpha P_R b)(\bar{\ell}\gamma^\alpha \ell), \notag \\
O_{10}^{\ell} &= (\bar{s}\gamma_\alpha P_L b)(\bar{\ell}\gamma^\alpha \gamma_5 \ell), \qquad &&O_{10}^{\prime \ell} = (\bar{s}\gamma_\alpha P_R b)(\bar{\ell}\gamma^\alpha \gamma_5 \ell).
\end{alignat}
We find it convenient to switch to the basis described in Ref.\ \cite{Hiller:2014yaa} where the effective Hamiltonian contains
\begin{align}
\mathcal{H}_{\text{eff}} \supset -\frac{4G_F}{\sqrt{2}}V_{tb}V^*_{ts}\frac{\alpha}{4\pi} \sum_{\ell = e,\mu} C_{LL}^\ell O_{LL}^{\ell} + \text{h.c.},
\end{align}
where $O^\ell_{LL} = (O_9^\ell - O_{10}^\ell)/2 = (\bar{s}\gamma_\alpha P_L b)(\bar{\ell}\gamma^\alpha P_{L} \ell)$ and $C_{LL}^\ell = C_9^\ell - C_{10}^\ell$, as well as the analogous operators with the other possible chiral structures. One potential way to explain the anomalies in $b \rightarrow s \mu \mu$ is to generate a large, in absolute value, and negative contribution to $C^{\mu}_{LL}$.\footnote{Below we often refer to generating large $C^{\mu}_{LL}$. By this we mean large in absolute value and negative.} Using all relevant data, the model independent analysis performed by Ref.\ \cite{Capdevila:2017bsm} finds the best fit value for $C^{\mu}_{LL}$ (assuming that only this coupling receives new physics contributions) to be $-1.24$ with the $2 \sigma$ range being $-1.76<C^{\mu}_{LL}< -0.74$.

In attempting to explain these anomalies, we consider the $R$-parity violating superpotential term $\lambda'_{ijk} L_i Q_j D^c_k$. In this expression, the $\lambda'$ couplings and the superfields are in a basis where the down-type quark mass matrix is diagonal. To switch to the mass basis, we assume that the scalar soft masses are diagonal in flavour space and apply a rotation to the left-handed up type superfields. Then, after expanding the superfields in terms of their fermions and sfermions, we get
\begin{align}
\mathcal{L} \supset &-\lambda'_{ijk}(\tilde{\nu}_i d_{Lj} \bar{d}_{Lk}  + \tilde{d}_{Lj} \nu_i \bar{d}_{Lk} + \tilde{d}_{Rk}^* \nu_i  d_{Lj}) \notag \\
&+\tilde{\lambda}'_{ijk}(\tilde{e}_{Li} u_{Lj} \bar{d}_{Lk} + \tilde{u}_{Lj} e_{Li}\bar{d}_{Lk} + \tilde{d}_{Rk}^* e_{Li}u_{Lj} ) + \text{h.c.,}
\end{align}
where we use 2-spinor notation to denote the fermion fields. In this equation, and throughout the rest of the paper unless otherwise stated, all repeated indices are assumed to be summed over. We have labeled the couplings involving left-handed down quarks and squarks as $\lambda'$ and the couplings involving left-handed up quarks and squarks as $\tilde{\lambda}'$. The $\lambda'$ and $\tilde{\lambda}'$ couplings are related by
\begin{align}
\tilde{\lambda}'_{ijk} = \lambda'_{ilk}V_{jl}^*.
\end{align}

\begin{figure}[t!]
  \centering
  \includegraphics[width=0.47\textwidth, bb = 0 0 319 170 ]{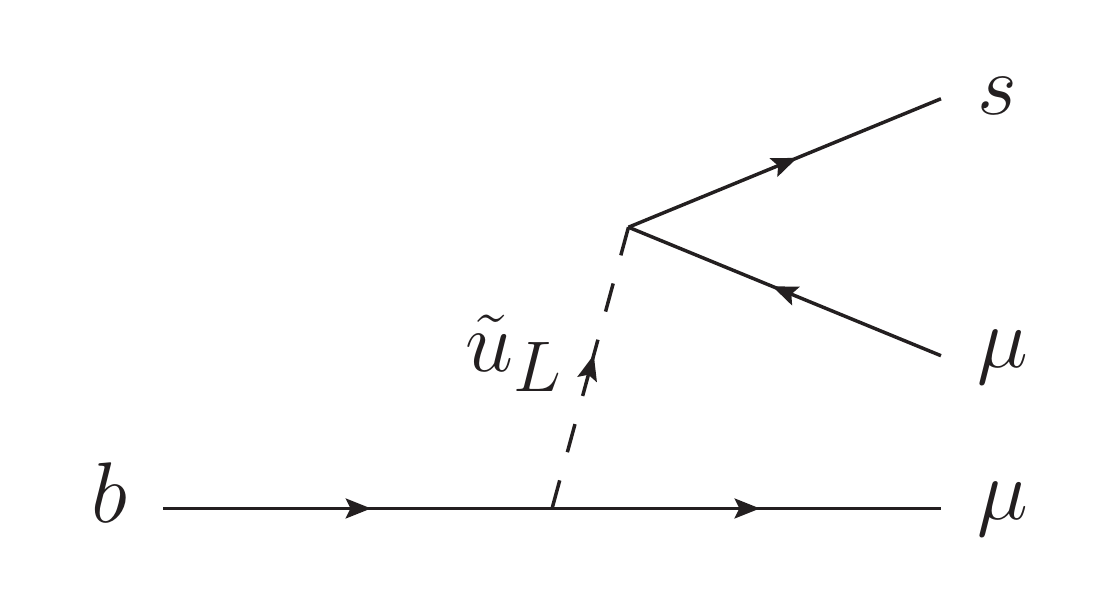}
  \caption{Tree level decay for $b\rightarrow s \mu\mu$ involving two $\lambda'$ interactions. }
  \label{fig:bsmumu_tree}
\end{figure}

As shown in figure \ref{fig:bsmumu_tree}, the decay $b \rightarrow s \mu \mu$ can occur at tree level through two $\lambda'$ interactions. After integrating out the left-handed up squark we are left with the effective Lagrangian
\begin{align}
\mathcal{L}_\text{eff} = -\frac{\tilde{\lambda}'_{2j2} \tilde{\lambda}'^*_{2j3}}{2 m_{\tilde{u}_{Lj}}^2} (\bar{s}\gamma^\alpha P_R b)(\bar{\mu}\gamma_\alpha P_L \mu) + \text{h.c.}
\end{align}
Notice that this tree level decay necessarily involves a right-handed quark current, and operators involving a right-handed quark current are unable to explain the anomalies. Since we are considering a spectrum which features left-handed up squarks, it is imperative to forbid these diagrams. To do so, we only consider non-zero $\lambda'_{ijk}$ for a single value of $k$. This is the same approach as taken in \cite{Das:2017kfo}. As will be discussed in section \ref{sec:tau_decays}, the couplings with $k = 1$ or $k = 2$ are excluded in the setup we consider due to $\tau$ decays. However, for the sake of generality, we choose to keep $k$ as a free index in the equations presented in this section. Accordingly, in these equations, the index $k$ is not assumed to be summed over.

\begin{figure}[t!]
  \centering
  \begin{subfigure}[b]{0.47\textwidth}
    \centering
    \includegraphics[width=\textwidth, bb = 0 0 429 218 ]{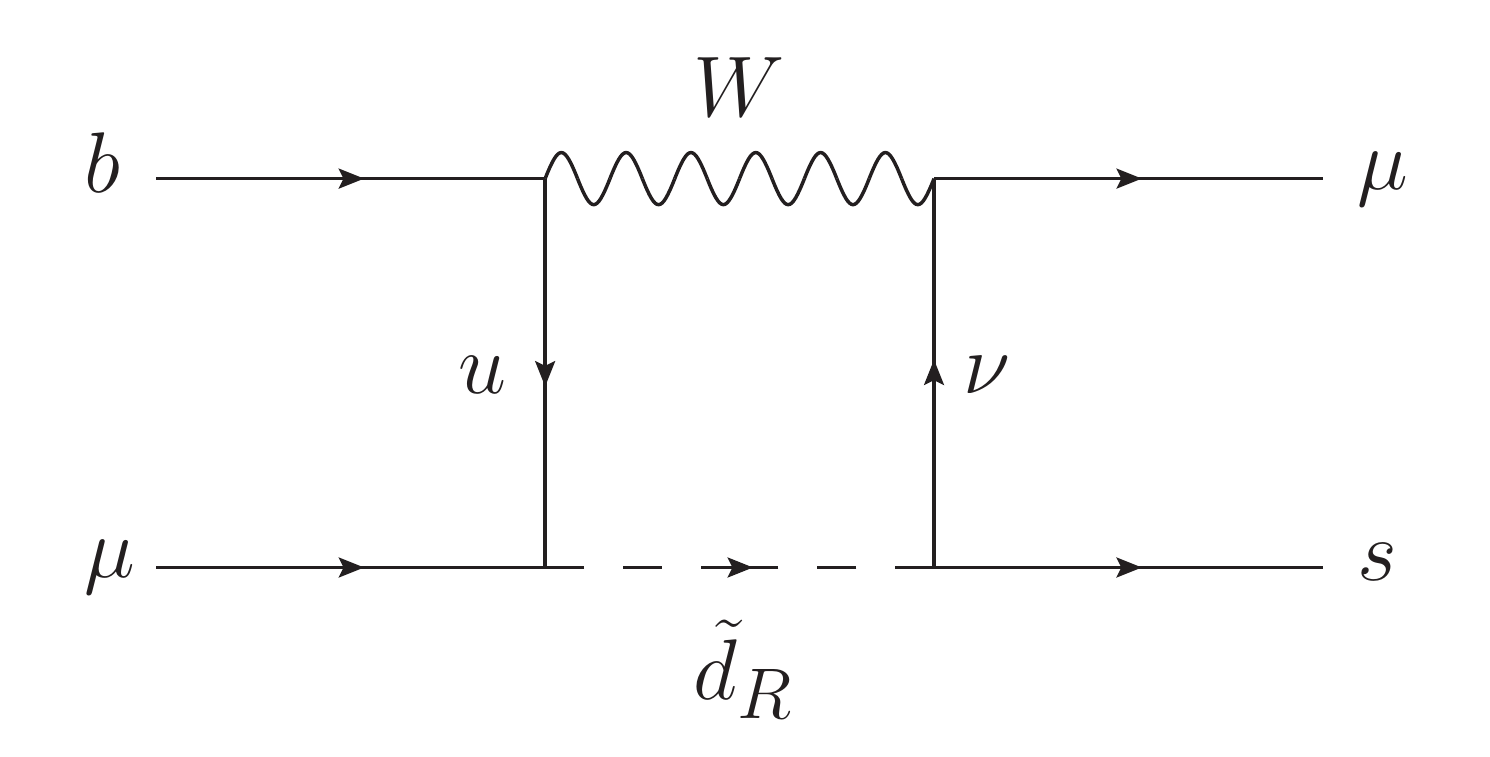}
    \caption{}
    \label{fig:W_loop}
  \end{subfigure}
  ~
  \begin{subfigure}[b]{0.47\textwidth}
    \centering
    \includegraphics[width=\textwidth, bb = 0 0 429 218 ]{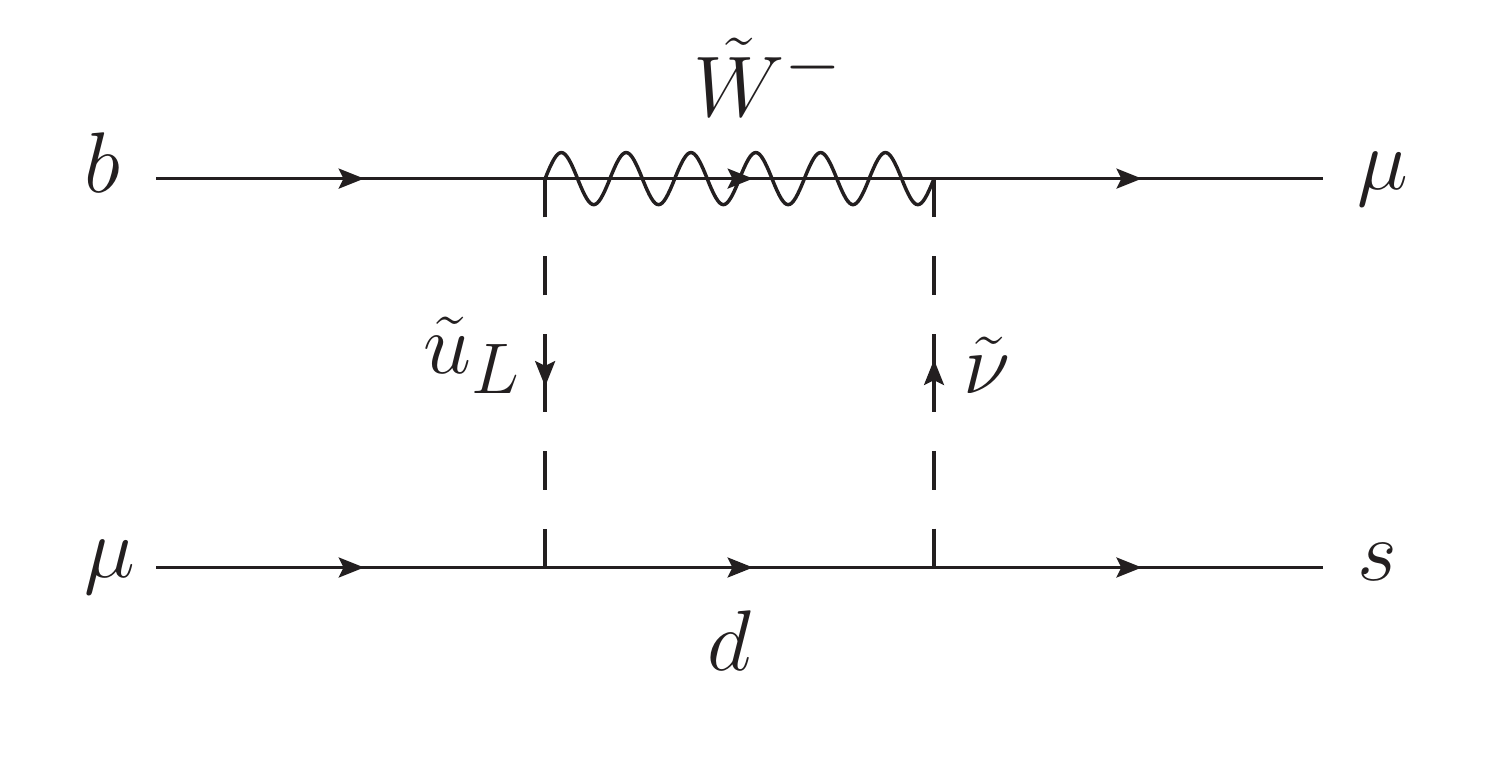}
    \caption{}
    \label{fig:wino_loop}
  \end{subfigure}
  ~
  \begin{subfigure}[b]{0.47\textwidth}
    \centering
    \includegraphics[width=\textwidth, bb = 0 0 429 218 ]{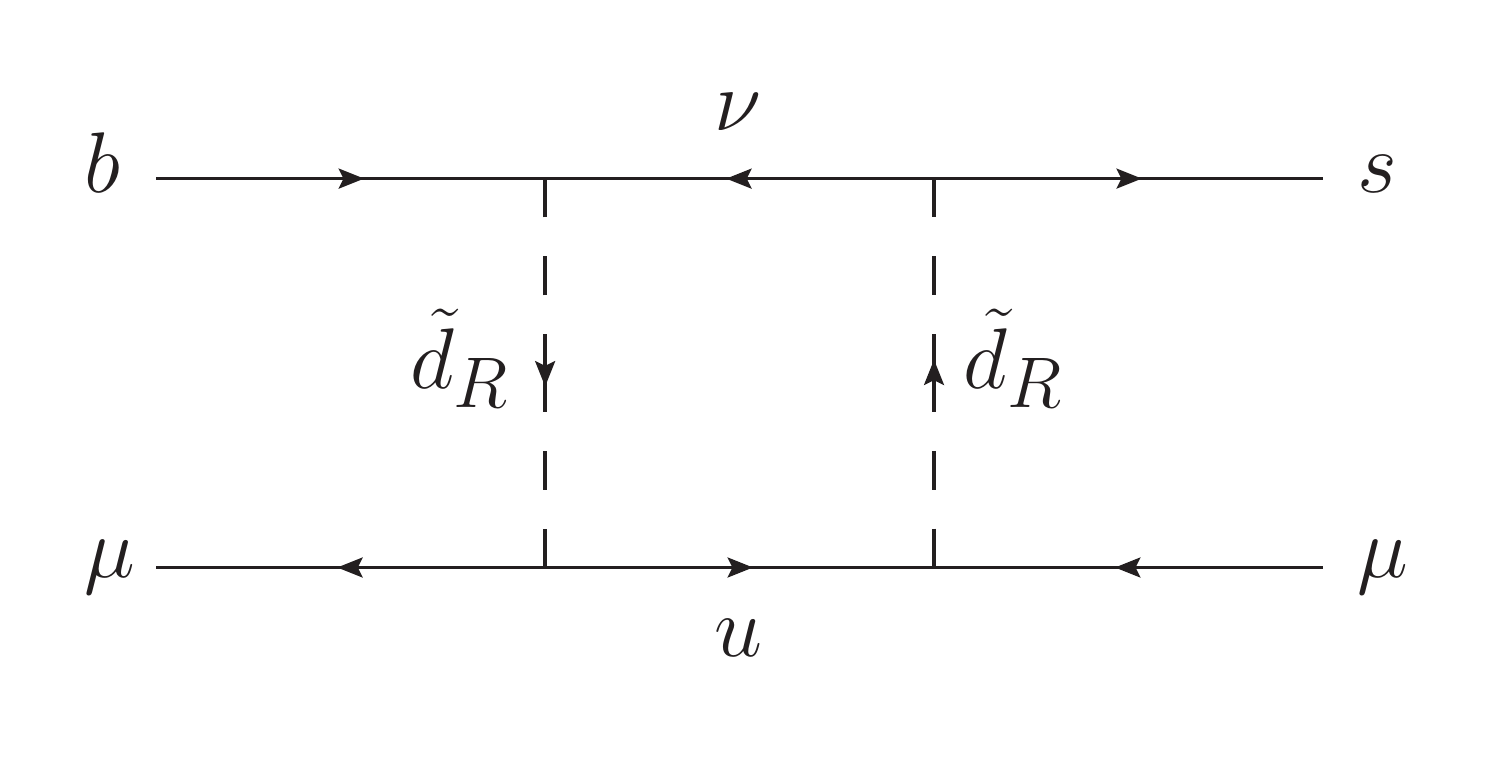}
    \caption{}
    \label{fig:TS_loop_1}
  \end{subfigure}
   ~
  \begin{subfigure}[b]{0.47\textwidth}
    \centering
    \includegraphics[width=\textwidth, bb = 0 0 429 218 ]{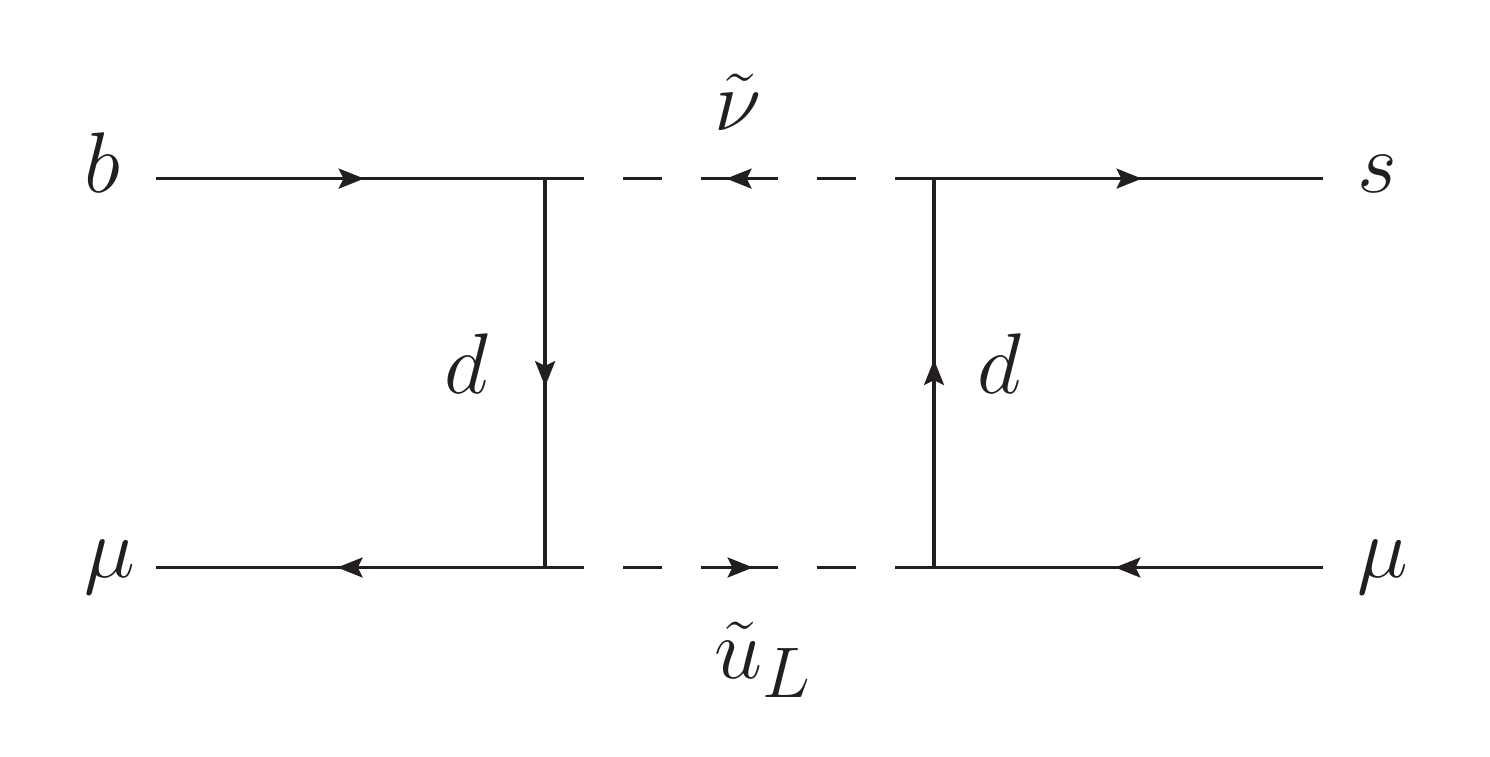}
    \caption{}
    \label{fig:TS_loop_2}
  \end{subfigure}
  \caption{Box diagrams studied in this work. Figure \ref{fig:W_loop} shows an example $W$ loop diagram, figure \ref{fig:wino_loop} shows an example wino loop diagram, and figures \ref{fig:TS_loop_1} and \ref{fig:TS_loop_2} show the four-$\lambda'$ loop diagrams. }\label{fig:bsmumu_loop}
\end{figure}

With the tree level decay forbidden, the next step is to examine potential loop level processes capable of mediating $b \rightarrow s \mu \mu$. Examples of the different box diagrams that we consider in this work are shown in figure \ref{fig:bsmumu_loop}.\footnote{It is worth noting that there are other potential one loop box diagrams for $b\rightarrow s \mu \mu$ involving $\lambda'$ and gauge couplings. However, these diagrams necessarily require the external quarks to be right-handed and thus, after Fierz rearrangements, will generate operators involving a right-handed quark current. Analogous to the tree level diagram, this is undesirable as operators involving a right-handed quark current are unable explain the anomalies. Fortunately, the same trick employed to forbid the tree level diagram, only turning on $\lambda'_{ijk}$ for a single value $k$, removes these diagrams as well.} First, consider the diagram involving a $W$ boson and a right-handed down squark, figure \ref{fig:W_loop}. This diagram is just one of many diagrams involving these two types of particles (if we ignore internal indices then there are four other diagrams, three with a $W$ boson and one with a Goldstone boson). Collectively, we refer to these diagrams as the $W$ loop diagrams. Second, consider the diagram involving a wino and a down quark, figure \ref{fig:wino_loop}. This diagram is just one of many diagrams involving these two types of particles (if we ignore internal indices then there are three other diagrams). Collectively, we refer to these diagrams as the wino loop diagrams. Finally, consider the diagrams involving four $\lambda'$ couplings, figures \ref{fig:TS_loop_1} and \ref{fig:TS_loop_2}. Collectively, we refer to these diagrams as the four-$\lambda'$ loop diagrams.

Each of these diagrams contribute to $C_{LL}^\mu$. Indeed, the $W$ loop diagrams and the four-$\lambda'$ loop diagrams have previously been considered in the literature in the context of the $b \rightarrow s \mu \mu$ anomalies. For example, Ref.\ \cite{Bauer:2015knc} studied a leptoquark model where equivalent diagrams to the $W$ loop and the four-$\lambda'$ loop with two right-handed down squarks were considered. Additionally, Ref.\ \cite{Das:2017kfo} studied an RPV supersymmetry model where the $W$ loop and both four-$\lambda'$ loop diagrams were considered. To the best of our knowledge, the wino loop diagrams have not been considered in the context of these anomalies. We now proceed by writing down the contributions of each of these diagrams to $C_{LL}^\mu$. Although the results for the $W$ loop diagrams and the four-$\lambda'$ loop diagrams can be found in the given references, we present them here for completeness.

First, it is convenient to introduce the integrals
\begin{align}
&D_0[m_1^2,m_2^2,m_3^2,m_4^2] \equiv \int \frac{d^4 k}{(2\pi)^4}\frac{1}{(k^2-m_1^2)(k^2-m_2^2)(k^2-m_3^2)(k^2-m_4^2)} \notag \\
&= -\frac{i}{16\pi^2} \biggl(\frac{m_1^2 \log(m_1^2)}{(m_1^2-m_2^2)(m_1^2-m_3^2)(m_1^2-m_4^2)} + (m_1 \leftrightarrow m_2) + (m_1 \leftrightarrow m_3) + (m_1 \leftrightarrow m_4)\biggr)
\end{align}
and 
\begin{align}
&D_2[m_1^2,m_2^2,m_3^2,m_4^2] \equiv \int \frac{d^4 k}{(2\pi)^4}\frac{k^2}{(k^2-m_1^2)(k^2-m_2^2)(k^2-m_3^2)(k^2-m_4^2)} \notag \\
&= -\frac{i}{16\pi^2} \biggl(\frac{m_1^4 \log(m_1^2)}{(m_1^2-m_2^2)(m_1^2-m_3^2)(m_1^2-m_4^2)} + (m_1 \leftrightarrow m_2) + (m_1 \leftrightarrow m_3) + (m_1 \leftrightarrow m_4)\biggr)
\end{align}
which arise when computing the box diagrams. These are simply the four-point Passarino-Veltman functions where the external momenta have been ignored \cite{Passarino:1978jh, Denner:1991kt}. These two integrals can also be written so that the arguments of the logarithms are dimensionless ratios of squared masses. We write them in this form to show the symmetry between $m_1^2$, $m_2^2$, $m_3^2$, and $m_4^2$. These integrals also have many well-defined limits when, for example, any of the masses are set to zero or any two masses are set equal. We will often use some of these limits below.

The contribution to $C_{LL}^\mu$ due to the $W$ loop diagrams is given by
\begin{align}
C_{LL}^{\mu(W)} = &\frac{\sqrt{2}}{4G_F}\frac{4\pi}{\alpha}\frac{1}{V_{tb}V_{ts}^*}\frac{1}{i}\biggl(\frac{g^2}{4} \tilde{\lambda}'_{2ik}\lambda'^*_{22k}V_{ib} D_2[m^2_{\tilde{d}_{Rk}},m^2_{u_i},m^2_{W},0] \notag \\
&- \frac{g^2}{4} \tilde{\lambda}'_{2ik}\tilde{\lambda}'^*_{2jk}V_{ib}V^*_{js} D_2[m^2_{\tilde{d}_{Rk}},m^2_{u_i},m^2_{u_j},m^2_W] + \frac{g^2}{4} \lambda'_{23k}\tilde{\lambda}'^*_{2jk}V^*_{js} D_2[m^2_{\tilde{d}_{Rk}},m^2_{u_j},m^2_{W},0] \notag \\ 
&- \frac{g^2}{4} \lambda'_{23k}\lambda'^*_{22k} D_2[m^2_{\tilde{d}_{Rk}},m^2_W,0,0]  + \tilde{\lambda}'_{2ik}\tilde{\lambda}'^*_{2jk}V_{ib}V^*_{js} \frac{m^2_{u_i}m^2_{u_j}}{2v^2} D_0[m^2_{\tilde{d}_{Rk}},m^2_{u_i},m^2_{u_j},m^2_W] \biggr)
\end{align}
where $v \approx 174 \ \text{GeV}$ is the vacuum expectation value of the Standard Model Higgs doublet. In the limit $m^2_{\tilde{d}_{Rk}} \gg m^2_t$, this simplifies to
\begin{align} \label{eq:CLLmuW}
C_{LL}^{\mu(W)} = \frac{|\lambda'_{23k}|^2}{8 \pi \alpha} \biggl( \frac{m_t^2}{m_{\tilde{d}_{Rk}}^2} \biggr).
\end{align}
Other combinations of $\lambda'$ couplings also contribute to $C_{LL}^{\mu(W)}$ but these are all much smaller. Next, the contribution from the wino loop diagrams is given by the similar expression
\begin{align}
C_{LL}^{\mu(\tilde{W})}& = \frac{\sqrt{2}}{4G_F}\frac{4\pi}{\alpha}\frac{1}{V_{tb}V_{ts}^*}\frac{1}{i}\biggl(\frac{g^2}{4} \tilde{\lambda}'_{2ik}\lambda'^*_{22k}V_{ib} D_2[m^2_{\tilde{W}},m^2_{\tilde{u}_{Li}},m^2_{\tilde{\nu}_{\mu}},m^2_{d_k}] \notag \\
&- \frac{g^2}{4} \tilde{\lambda}'_{2ik}\tilde{\lambda}'^*_{2jk}V_{ib}V^*_{js} D_2[m^2_{\tilde{W}},m^2_{\tilde{u}_{Li}},m^2_{\tilde{u}_{Lj}},m^2_{d_k}] + \frac{g^2}{4} \lambda'_{23k}\tilde{\lambda}'^*_{2jk}V^*_{js} D_2[m^2_{\tilde{W}},m^2_{\tilde{u}_{Lj}},m^2_{\tilde{\nu}_{\mu}},m^2_{d_k}] \notag \\ 
& - \frac{g^2}{4} \lambda'_{23k}\lambda'^*_{22k} D_2[m^2_{\tilde{W}},m^2_{\tilde{\nu}_{\mu}},m^2_{\tilde{\nu}_{\mu}},m^2_{d_k}] \biggr).
\end{align}
If we make the assumption that the masses of the three left-handed up squarks are degenerate, then this simplifies to
\begin{alignat}{2} \label{eq:CLLmuWino}
C_{LL}^{\mu(\tilde{W})} &= &&\frac{\sqrt{2}g^2 \lambda'_{23k}\lambda'^*_{22k}}{64\pi G_F \alpha V_{tb}V^*_{ts}m^2_{\tilde{W}}} \biggl(\frac{1}{x_{\tilde{\nu}_\mu} - 1} + \frac{1}{x_{\tilde{u}_L} - 1} \notag \\
& && + \frac{(x_{\tilde{\nu}_\mu} - 2 x^2_{\tilde{\nu}_\mu} + x_{\tilde{u}_L})\log(x_{\tilde{\nu}_\mu})}{(x_{\tilde{\nu}_\mu} - 1)^2(x_{\tilde{\nu}_\mu} - x_{\tilde{u}_L})} + \frac{(x_{\tilde{u}_L} - 2 x^2_{\tilde{u}_L} + x_{\tilde{\nu}_\mu})\log(x_{\tilde{u}_L})}{(x_{\tilde{u}_L} - 1)^2(x_{\tilde{u}_L} - x_{\tilde{\nu}_\mu})} \biggr)
\end{alignat}
where $x_{\tilde{\nu}_\mu} = m^2_{\tilde{\nu}_{\mu}}/m^2_{\tilde{W}}$, $x_{\tilde{u}_L} = m^2_{\tilde{u}_{L}}/m^2_{\tilde{W}}$, and we have set $m_{d_k}^2 \rightarrow 0$. Notice that if $x_{\tilde{\nu}_\mu} = x_{\tilde{u}_L}$, then this expression vanishes due to a super GIM mechanism. Another relevant limit is $x_{\tilde{\nu}_\mu} \gg x_{\tilde{u}_L}$, in which case $C_{LL}^{\mu(\tilde{W})}$ further simplifies to
\begin{align} \label{eq:CLLmuWino2}
C_{LL}^{\mu(\tilde{W})} &= \frac{\sqrt{2}g^2 \lambda'_{23k}\lambda'^*_{22k}}{64\pi G_F \alpha V_{tb}V^*_{ts}m^2_{\tilde{W}}}\biggl(\frac{1}{x_{\tilde{u}_L} - 1} - \frac{\log(x_{\tilde{u}_L})}{(x_{\tilde{u}_L} - 1)^2} \biggr)
\end{align}
which is simply the result of the box diagram with two left-handed up squarks in the loop. Finally, the contribution from the four-$\lambda'$ loop diagrams is given by
\begin{align}
C_{LL}^{\mu(4\lambda')} = \frac{\sqrt{2}}{4G_F}\frac{4\pi}{\alpha}\frac{1}{V_{tb}V_{ts}^*}\frac{1}{4} \lambda'_{i3k} \lambda'^*_{i2k} \tilde{\lambda}'_{2jk} \tilde{\lambda}'^*_{2jk} \frac{1}{i}\biggl(&D_2[m^2_{\tilde{d}_{Rk}},m^2_{\tilde{d}_{Rk}},m^2_{u_j},0] \notag \\
+ &D_2[m^2_{\tilde{u}_{Lj}},m^2_{\tilde{\nu}_i},m^2_{d_k},m^2_{d_k}]\biggr).
\end{align}
Assuming the masses of the three left-handed up squarks are degenerate and taking the limit $m^2_{\tilde{d}_{Rk}} \gg m^2_t$, this simplifies to
\begin{align} \label{eq:CLLmu4lambda}
C_{LL}^{\mu(4\lambda')} = -\frac{\sqrt{2}\lambda'_{i3k} \lambda'^*_{i2k} \lambda'_{2jk} \lambda'^*_{2jk}}{64\pi G_F \alpha V_{tb}V^*_{ts}}\biggl(\frac{1}{m^2_{\tilde{d}_{Rk}}} + \frac{\log(m^2_{\tilde{\nu}_i}/m^2_{\tilde{u}_{L}})}{m^2_{\tilde{\nu}_i} - m^2_{\tilde{u}_{L}}} \biggr).
\end{align}

\begin{figure}[t!]
  \centering
  \includegraphics[width=0.47\textwidth, bb = 0 0 319 239 ]{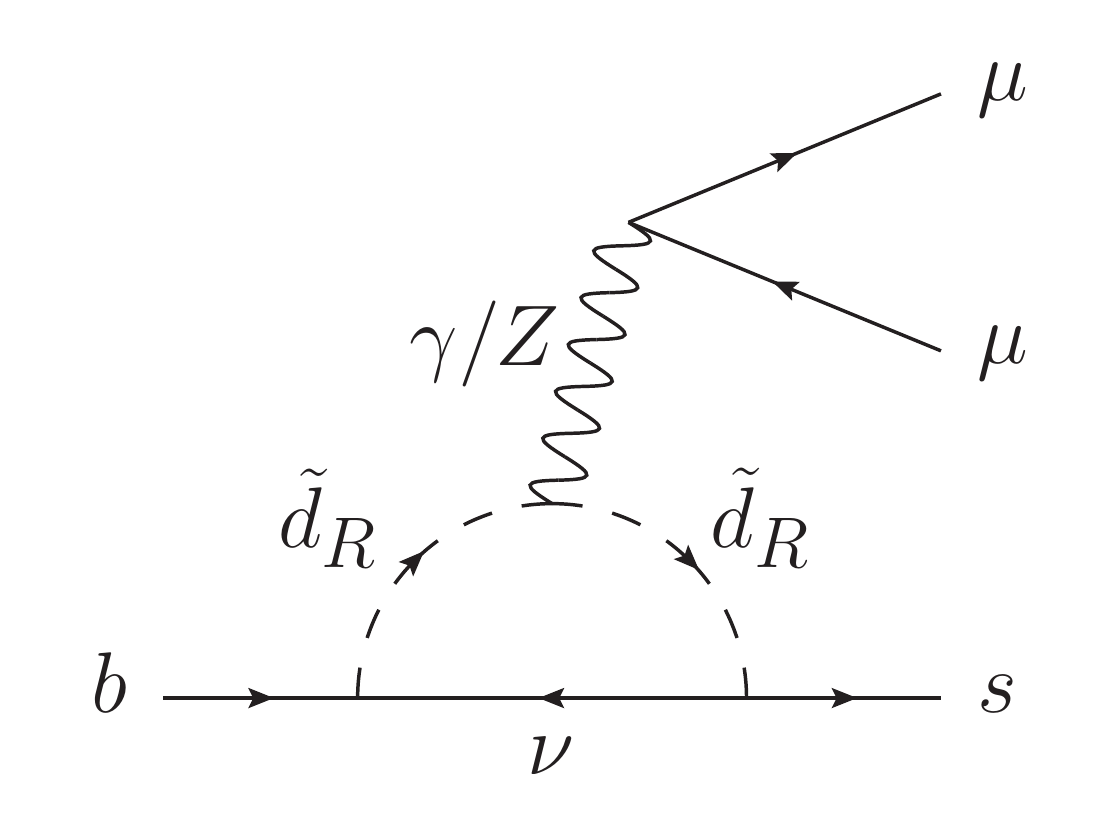}
  \caption{An example penguin diagram for $b\rightarrow s \mu\mu$.}
  \label{fig:penguin}
\end{figure}

So far, we have considered only box diagrams for $b \rightarrow s \mu \mu$. We now consider potential photonic and $Z$ penguin contributions, for which an example diagram is shown in figure \ref{fig:penguin}. Starting with the photonic penguin, we determine its contribution to $C_{LL}^\mu$ as follows. Consider first the generic amplitude for the process $\bar{b} \rightarrow \bar{s} \gamma^{(*)}$
\begin{align} \label{eq:bsgamma}
i\mathcal{M} = i e \epsilon^{\alpha *} \bar{v}_{b}(p)[\gamma^\beta(g_{\alpha \beta}q^2 - q_{\alpha}q_{\beta})(A_{b1}^L P_L + A_{b1}^R P_R) + m_b \sigma_{\alpha \beta} i q^\beta(A_{b2}^L P_L + A_{b2}^R P_R)]v_{s}(p-q).
\end{align}
Adapting the results of Ref.\ \cite{deGouvea:2000cf}, who study the process $\mu^+ \rightarrow e^+ \gamma^{(*)}$ with $R$-parity violation, we find
\begin{align}
A_{b1}^L &= \frac{1}{3} \frac{\lambda'_{i23}\lambda'^*_{i33}}{16\pi^2}\biggl(-\frac{1}{3}\biggl(\frac{4}{3} + \log\biggl(\frac{m_b^2}{m_{\tilde{\nu}_i}^2} \biggr) \biggr)\frac{1}{m_{\tilde{\nu}_i}^2} + \frac{1}{18 m_{\tilde{b}_R}^2} \biggr), \label{eq:bsgammaA1L} \\
A_{b1}^R &= 0,
\end{align}
as well as
\begin{align}
A_{b2}^L &= \frac{1}{3}\frac{\lambda'_{i23}\lambda'^*_{i33}}{16\pi^2} \biggl(\frac{1}{12 m_{\tilde{b}_R}^2} - \frac{1}{6 m_{\tilde{\nu}_i}^2} \biggr), \label{eq:bsgammaA2L} \\
A_{b2}^R &= 0,
\end{align}
where we have momentarily considered the case $k = 3$. Here, $A_{b1}^R$ and $A_{b2}^R$ are zero because we are only considering non-zero $\lambda'_{ijk}$ couplings for a single value of $k$. Next, we match this amplitude onto effective operators. To resolve any potential sign ambiguities, we compare the effective operator for the dipole term with the results present in the literature \cite{deCarlos:1996yh, Besmer:2000rj}. From these effective operators, we determine a photonic penguin contribution to $C_{LL}^\mu$ given by
\begin{align}
C_{LL}^{\mu(\gamma)} = - \frac{\sqrt{2}\lambda'_{i33}\lambda'^*_{i23}}{12 G_F V_{tb}V^*_{ts}}\biggl(-\frac{1}{3}\biggl(\frac{4}{3} + \log\biggl(\frac{m_b^2}{m_{\tilde{\nu}_i}^2} \biggr) \biggr)\frac{1}{m_{\tilde{\nu}_i}^2} + \frac{1}{18 m_{\tilde{b}_R}^2} \biggr), \label{eq:CLLmugamma}
\end{align}
as well as an equal contribution to $C_{LR}^\mu$ as defined in \cite{Hiller:2014yaa}. Notice, however, that $C_{LL}^e$ and $C_{LR}^e$ will receive identical contributions. Thus the photonic penguin diagrams should not have any effect on lepton universality violating observables such as $R_{K^{(*)}}$. On the other hand, they should still affect the other types of variables, such as the various angular observables, used when making the fits for the Wilson coefficients. Regardless, it so happens that, in the setup we consider, all potential contributions from the photonic penguin diagrams are very small. We decide to add $C_{LL}^{\mu(\gamma)}$ to $C_{LL}^\mu$ but emphasize that this only has a negligible effect. Finally, we find that the $Z$ penguin diagrams vanish in the limit of zero down-type quark masses.

To explain the anomalies, we need to generate negative contributions to $C_{LL}^\mu$. From equation \ref{eq:CLLmuW}, we see that the $W$ loop diagrams necessarily give a positive contribution. Next, the term in the large brackets in equation \ref{eq:CLLmuWino} is positive for all values of $x_{\tilde{\nu}_\mu}$ and $x_{\tilde{u}_L}$. Assuming real $\lambda'$, which we do for the remainder of this section, we need to take the product $\lambda'_{22k}\lambda'_{23k} > 0$ to make $C_{LL}^{\mu(\tilde{W})}$ negative. Further, as previously mentioned, $C_{LL}^{\mu(\tilde{W})}$ vanishes in the limit $x_{\tilde{\nu}_\mu} = x_{\tilde{u}_L}$ due to a super GIM mechanism. As a result, to increase the magnitude of $C_{LL}^{\mu(\tilde{W})}$ it is beneficial to split the muon sneutrino and left-handed up squark masses. Taking the muon sneutrino mass much larger than the left-handed up squark masses leads to equation \ref{eq:CLLmuWino2}. Finally, by examining equation \ref{eq:CLLmu4lambda}, we see that if $\lambda'_{22k}\lambda'_{23k} > 0$, then $C_{LL}^{\mu(4\lambda')}$ receives a positive contribution. On the other hand, if we take $\lambda'_{12k}\lambda'_{13k} < 0$ or $\lambda'_{32k}\lambda'_{33k} < 0$ then this will result in negative contributions to $C_{LL}^{\mu(4\lambda')}$.

With these considerations, we envision the following spectrum. The masses of the wino and the three left-handed up squarks are light, of order $1 \ \text{TeV}$. The product $\lambda'_{22k}\lambda'_{23k}$ is positive and, to enhance the wino loop diagrams, fairly large. As we will see in section \ref{sec:B_constraints}, the product $\lambda'_{22k}\lambda'_{23k}$ is highly constrained by $B_s - \bar{B}_s$ mixing. To get around this constraint, the sfermions which enable $B_s - \bar{B}_s$ mixing with $\lambda'$ interactions, the right-handed down squarks and sneutrinos, must be made heavy. We set the masses of these particles to order $10 \ \text{TeV}$. The $W$ loop diagrams and the four-$\lambda'$ loop diagrams proportional to $\lambda'_{22k}\lambda'_{23k}$, which each give positive contributions to $C_{LL}^\mu$, are then suppressed. Furthermore, we find that it is still difficult to generate large enough $C_{LL}^\mu$ to explain the anomalies in this setup. Thus, we also turn on the product $\lambda'_{32k}\lambda'_{33k}$ and make it negative so that the four-$\lambda'$ loop diagrams proportional to this product of couplings then give negative contributions to $C_{LL}^\mu$. In fact, if we take $-\lambda'_{32k} \lambda'_{33k} > \lambda'_{22k} \lambda'_{23k}$ then $C_{LL}^{\mu(4\lambda')}$ will be negative. However, we must then consider constraints involving taus. One such constraint, examined in section \ref{sec:tau_decays}, is $\tau$ decays to a $\mu$ and a meson. There we find that the cases $k = 1$ or $k = 2$ are ruled out, and we are forced to consider $k = 3$. Due to this, the only right-handed down squark which is now relevant is the sbottom. In summary, we consider a light wino, light left-handed up squarks, a heavy right-handed sbottom, heavy sneutrinos, and the four $R$-parity violating couplings $\lambda'_{223}$, $\lambda'_{233}$, $\lambda'_{323}$, $\lambda'_{333}$ with $\lambda'_{223}\lambda'_{233} > 0$ and $\lambda'_{323}\lambda'_{333} < 0$.

There are two last points we wish to make before discussing potential constraints. First, we have chosen to turn on $\lambda'_{323}\lambda'_{333}$ instead of $\lambda'_{123}\lambda'_{133}$. There are two reasons for making this choice. The first is that if $\lambda'_{123}\lambda'_{133}$ is taken to be non-zero, then there will be diagrams contributing to $C_{LL}^e$. We avoid this since the fits, using all relevant observables, tend to prefer new physics in the muon channel than in the electron channel. Interpreting our results would also become much more challenging. The second reason for this choice of parameters is that by turning on $\lambda'_{323}\lambda'_{333}$ instead of $\lambda'_{123}\lambda'_{133}$ we need only to consider weaker constraints involving taus as opposed to stronger constraints involving electrons. For example, in section \ref{sec:tau_decays} we consider constraints from $\tau \rightarrow \mu\mu\mu$. This process is much less constrained than $\mu \rightarrow eee$. Finally, the last point we make is that taking $\lambda'_{223}\lambda'_{233} > 0$ and $\lambda'_{323}\lambda'_{333} < 0$ has an additional benefit, it tends to cause cancellations amongst diagrams contributing to potentially constraining processes. For example, as we will see in section \ref{sec:B_constraints}, such cancellations happen in $B_s - \bar{B}_s$ mixing. We consider these cancellations a feature of the model, as the choice of parameters which lead to them is what is precisely preferred to explain the anomalies.

\section{Constraints}\label{sec:constraints}

\subsection{$\tau$ decays}\label{sec:tau_decays}

The first type of constraints we discuss are those which follow from $\tau$ decays to a $\mu$ and a meson. This type of process was considered in \cite{Kim:1997rr} (see also \cite{Barbier:2004ez}) to bound various combinations of RPV couplings. We will show the results in \cite{Kim:1997rr} which are relevant to our parameter space and update the bounds using the latest experimental data. 

This type of process can be divided into two subcategories, $\tau \rightarrow \mu V$ and $\tau \rightarrow \mu P$, where $V$ represents a vector meson and $P$ a pseudoscalar. Both types of $\tau$ decays can occur via a tree level exchange of a $\tilde{u}_L$ or a $\tilde{d}_R$ depending on which meson is in the final state. However, as also noted in \cite{Barbier:2004ez}, we find that stronger constraints come from $\tau$ decays to vector mesons than from $\tau$ decays to pseudoscalars. Particularly, the mesons which give the strongest bounds are $\rho^0$ and $\phi$. The branching ratio for the decay $\tau \rightarrow \mu V$ is given by \cite{Kim:1997rr}
\begin{align}
\text{Br}(\tau \rightarrow \mu V) = \frac{1}{512\pi} |A_V|^2 f_V^2 m_{\tau}^3 \biggl(1 + \frac{m_V^2}{m_\tau^2} - 2 \frac{m_V^4}{m_\tau^4}\biggr)\biggl(1 - \frac{m_V^2}{m_\tau^2}\biggr) \tau_{\tau},
\end{align}
where $\tau_\tau$ is the mean lifetime of the $\tau$ and we have taken the $m_{\mu}^2 / m_{\tau}^2 \rightarrow 0$ limit. The vector meson decay constant $f_V$ is defined by \cite{Kim:1997rr}
\begin{align}
\braket{\rho^0(p,\epsilon)|\bar{u}\gamma_\alpha u(0)|0} &= m_\rho f_\rho \epsilon_\alpha^* = -\braket{\rho^0(p,\epsilon)|\bar{d}\gamma_\alpha d(0)|0} 
\end{align} 
for $\rho^0$, and 
\begin{align}
\braket{\phi(p,\epsilon)|\bar{s}\gamma_\alpha s(0)|0} &= m_\phi f_\phi \epsilon_\alpha^*
\end{align} 
for $\phi$, with $f_\rho = 153 \ \text{MeV}$ and $f_\phi = 237\ \text{MeV}$. Additionally, $A_V$ is given by \cite{Kim:1997rr} 
\begin{align}
A_{\rho^0} = \frac{\tilde{\lambda}'_{3j1} \tilde{\lambda}'^{*}_{2j1}}{m^2_{\tilde{u}_{Lj}}} - \frac{\tilde{\lambda}'_{31k}\tilde{\lambda}'^{*}_{21k}}{m^2_{\tilde{d}_{Rk}}} 
\end{align}
for $\rho^0$, and 
\begin{align}
A_{\phi} = \frac{\tilde{\lambda}'_{3j2} \tilde{\lambda}'^{*}_{2j2}}{m_{\tilde{u}_{Lj}}^2}
\end{align}
for $\phi$. The current experimental upper limits on the branching ratios for these two processes are $\text{Br}(\tau \rightarrow \mu \rho^0) < 1.2 \times 10^{-8}$ and $\text{Br}(\tau \rightarrow \mu \phi) < 8.4 \times 10^{-8}$ \cite{Patrignani:2016xqp}. These translate into the bounds
\begin{align}
\biggl|\tilde{\lambda}'_{3j1} \tilde{\lambda}'^{*}_{2j1} \biggl(\frac{1\text{TeV}}{m_{\tilde{u}_{Lj}}}\biggr)^2 - \tilde{\lambda}'_{31k}\tilde{\lambda}'^{*}_{21k} \biggl(\frac{1\text{TeV}}{m_{\tilde{d}_{Rk}}} \biggr)^2 \biggr| < 0.019
\end{align}
and
\begin{align}
\biggl| \tilde{\lambda}'_{3j2} \tilde{\lambda}'^{*}_{2j2} \biggl(\frac{1\text{TeV}}{m_{\tilde{u}_{Lj}}}\biggr)^2 \biggr| < 0.036,
\end{align}
respectively. As we are considering the masses of the left-handed up squarks to be of order $1 \ \text{TeV}$, these two bounds are highly constraining. Indeed, explaining the anomalies with the couplings $\lambda'_{22k}$, $\lambda'_{23k}$, $\lambda'_{32k}$, and $\lambda'_{33k}$ with $k = 1$ or $k = 2$ proves to be impossible due to these stringent limits. This is why we are forced to consider the couplings $\lambda'_{223}$, $\lambda'_{233}$, $\lambda'_{323}$, and $\lambda'_{333}$. Below, we will discuss constraints which would otherwise depend on $m_{\tilde{d}_{Rk}}$. However, because of this restriction, we will only mention the right-handed sbottom from here on out. 

Other $\tau$ decays which can potentially constrain the parameter space include $\tau \rightarrow \mu \gamma$ and the similar processes $\tau \rightarrow \mu \mu \mu$ and $\tau \rightarrow \mu e^+ e^-$. The processes $\mu \rightarrow e \gamma$ and $\mu \rightarrow e e e$ in the context of RPV supersymmetry are considered in detail in Ref.\ \cite{deGouvea:2000cf} and we modify their results for $\tau$ decays. First, note that the amplitude for $\tau^+ \rightarrow \mu^+ \gamma^{(*)}$ is the same, up to appropriate modifications, as the amplitude given in equation \ref{eq:bsgamma}. The dipole term contributes to the decay $\tau \rightarrow \mu \gamma$ and leads to a branching ratio of \cite{deGouvea:2000cf}
\begin{align}
\text{Br}(\tau \rightarrow \mu \gamma) = \frac{\alpha m_\tau^5}{4}(|A_{\tau 2}^{L}|^2 + |A_{\tau 2}^{R}|^2) \tau_\tau
\end{align}
where we have again taken the $m_{\mu}^2 / m_{\tau}^2 \rightarrow 0$ limit, and \cite{deGouvea:2000cf}
\begin{align}
A_{\tau 2}^{L} &= - \frac{ {\lambda}'_{2j3} {\lambda}'^{*}_{3j3} }{64 \pi^2 m_{\tilde{b}_R}^2}, \label{eq:tauA2L} \\
A_{\tau 2}^R &= 0. \label{eq:tauA2R}
\end{align}
Interestingly, $A_{\tau 2}^L$ does not depend on the masses of the left-handed up squarks, even though there are diagrams which involve these particles. This is because in the limit $m^2_b / m^2_{\tilde{u}_L} \rightarrow 0$ and $m^2_\tau / m^2_{\tilde{u}_L} \rightarrow 0$ there is an exact cancellation amongst the individual diagrams. Also worth noting is that to reach $A_{\tau 2}^L$ shown above we have taken the $m_t^2/m^2_{\tilde{b}_R} \rightarrow 0$ limit. The branching ratio then becomes
\begin{align}
\text{Br}(\tau \rightarrow \mu \gamma) = \frac{\alpha m_\tau^5}{16384 \pi^4 m_{\tilde{b}_R}^4} |\lambda'_{223}\lambda'^{*}_{323}  + \lambda'_{233}\lambda'^{*}_{333} |^2 \tau_\tau
\end{align}
and this, using the current experimental upper limit $\text{Br}(\tau \rightarrow \mu \gamma) < 4.4 \times 10^{-8}$ \cite{Patrignani:2016xqp}, leads to the bound
\begin{align}
|\lambda'_{223}\lambda'^{*}_{323}  + \lambda'_{233}\lambda'^{*}_{333}| < 1.1 \biggl(\frac{m_{\tilde{b}_R}}{1\text{TeV}}\biggr)^2.
\end{align}
Since we are considering $m_{\tilde{b}_R}$ to be of order $10 \ \text{TeV}$, we find no constraints from $\tau \rightarrow \mu \gamma$.

\begin{figure}[t!]
  \centering
  \begin{subfigure}[b]{0.47\textwidth}
    \centering
    \includegraphics[width=\textwidth, bb = 0 0 434 209 ]{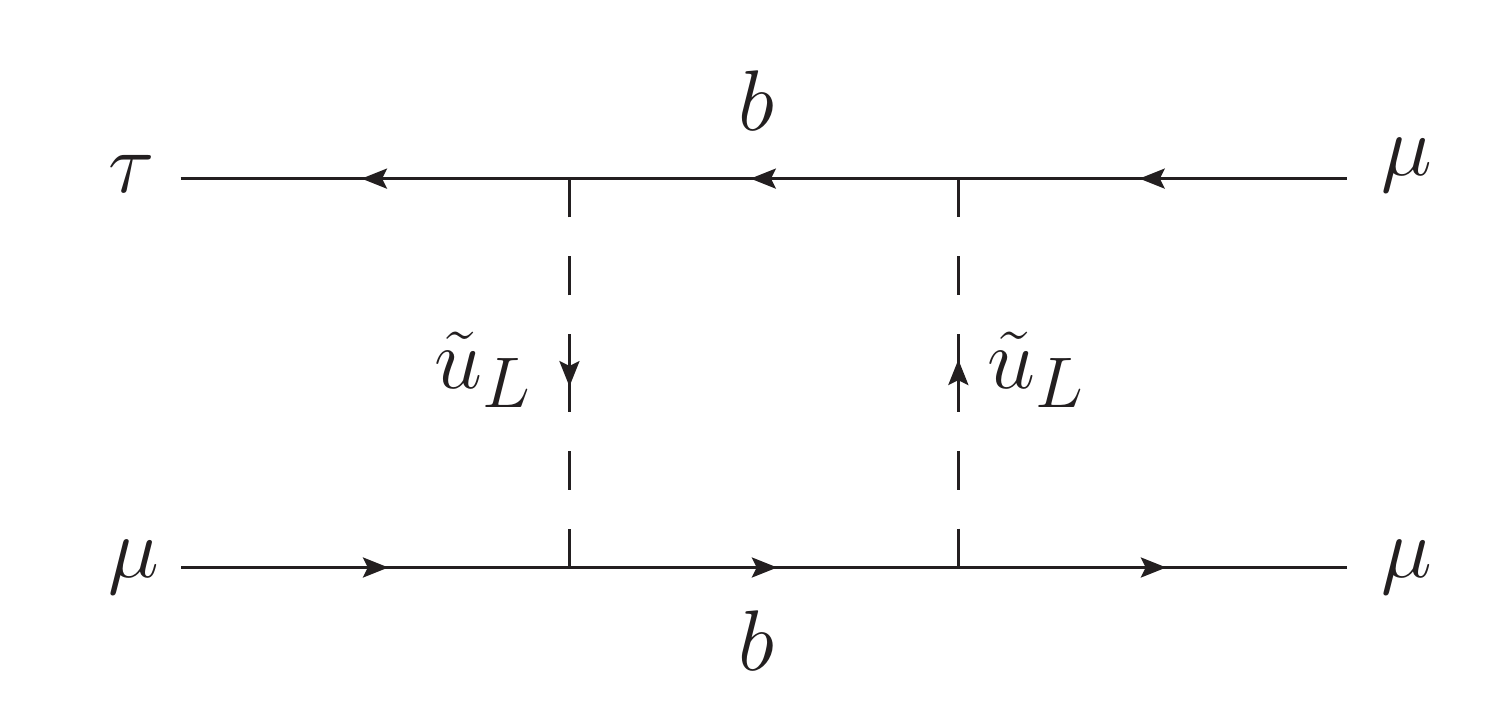}
    \caption{}
    \label{fig:tau_mumumu_1}
  \end{subfigure}
  ~
  \begin{subfigure}[b]{0.47\textwidth}
    \centering
    \includegraphics[width=\textwidth, bb = 0 0 434 209 ]{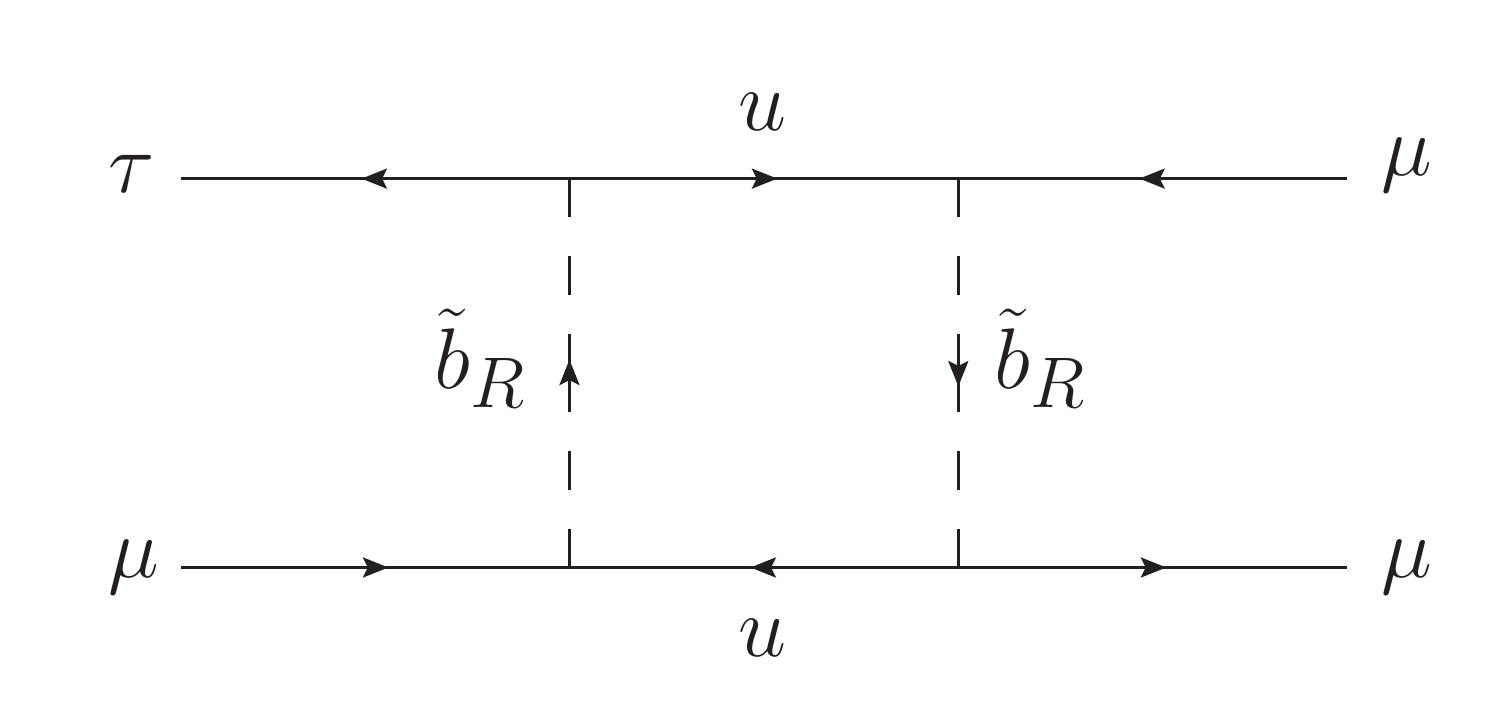}
    \caption{}
    \label{fig:tau_mumumu_2}
  \end{subfigure}
  \caption{One loop box diagrams contributing to $\tau \rightarrow \mu \mu \mu$.}\label{fig:tau_mumumu}
\end{figure}

Next we consider the decay $\tau^+ \rightarrow \mu^+ \mu^+ \mu^-$. This decay receives three different types of contributions, photonic and $Z$ penguin diagrams and box diagrams with four $\lambda'$ couplings. We write this as
\begin{align}
i \mathcal{M} = i\mathcal{M}^\gamma + i\mathcal{M}^Z + i\mathcal{M}^{4\lambda'}.
\end{align}
The photonic penguin amplitude $i\mathcal{M}^\gamma$ is given by
\begin{alignat}{2} \label{eq:taumumumu_photon}
i\mathcal{M}^\gamma &= i e^2 &&[\bar{v}_\tau(p)\biggl(\gamma_\alpha (A_{\tau 1}^L P_L + A_{\tau 1}^R P_R) + m_\tau \sigma_{\alpha\beta} \frac{iq^\beta}{q^2}(A_{\tau 2}^L P_L + A_{\tau 2}^R P_R) \biggr)v_\mu(p_2)] \notag \\
& &&[\bar{u}_\mu(p_3)\gamma^\alpha v_\mu(p_1)] - (p_1 \leftrightarrow p_2).
\end{alignat}
The functions $A_{\tau 2}^L$ and $A_{\tau 2}^R$ are still given by equations \ref{eq:tauA2L} and \ref{eq:tauA2R}, respectively, and $A_{\tau 1}^R = 0$. The function $A_{\tau 1}^L$ is similar in nature to equation \ref{eq:bsgammaA1L} but is slightly more complicated. Its exact form can be determined from the results in \cite{deGouvea:2000cf}. We do note though that, unlike $A_{\tau 2}^L$, $A_{\tau 1}^L$ does depend on the masses of the left-handed up squarks. Next, the amplitude for the $Z$ penguin is given by
\begin{align} \label{eq:taumumumu_Z}
i\mathcal{M}^{Z} = i \frac{g^2}{32 \pi^2 c_W^2 m_Z^2} B_{32}^2 [\bar{v}_{\tau}(p)\gamma^\alpha P_L v_{\mu}(p_2)][\bar{u}_{\mu}(p_3)\gamma_\alpha (\kappa_L P_L + \kappa_R P_R) v_{\mu}(p_1)] - (p_1 \leftrightarrow p_2)
\end{align}
where $\kappa_L = -\frac{1}{2} + s_W^2$, $\kappa_R = s_W^2$, $c_W = \cos\theta_W$, $s_W = \sin\theta_W$, and the function $B_{32}^2$ is given in equation \ref{eq:Zamp2} with $m_Z^2 \rightarrow 0$. Finally, consider the two box diagrams shown in figure \ref{fig:tau_mumumu}. The amplitude for these two diagrams is
\begin{align}
i\mathcal{M}^{4\lambda'} = i C_\tau [\bar{v}_\tau(p)\gamma^\alpha P_L v_\mu(p_2)][\bar{u}_\mu(p_3)\gamma_\alpha P_L v_\mu(p_1)] - (p_1 \leftrightarrow p_2)
\end{align}
where $C_\tau$ is given by
\begin{align}
C_\tau = -\frac{1}{4}\tilde{\lambda}'_{2i3}\tilde{\lambda}'^*_{2i3} \tilde{\lambda}'_{2j3}\tilde{\lambda}'^*_{3j3}\frac{1}{i}(D_2[m^2_{\tilde{u}_{Li}},m^2_{\tilde{u}_{Lj}},m_b^2,m_b^2] + D_2[m^2_{\tilde{b}_R},m^2_{\tilde{b}_R},m^2_{u_i},m^2_{u_j}] ).
\end{align}
Assuming mass degenerate left-handed up squarks, $m^2_{\tilde{u}_L} \gg m^2_b$, and $m^2_{\tilde{b}_R} \gg m^2_t$, this simplifies to
\begin{align}
C_\tau = \frac{\lambda'_{2i3}\lambda'^*_{2i3} \lambda'_{2j3}\lambda'^*_{3j3}}{64 \pi^2}\biggl(\frac{1}{m^2_{\tilde{u}_L}} + \frac{1}{m^2_{\tilde{b}_R}} \biggr).
\end{align}
To compute potential limits from $\tau \rightarrow \mu\mu\mu$, we first write the amplitude in Mathematica with the assistance of FeynCalc \cite{Mertig:1990an,Shtabovenko:2016sxi}. Then, also using FeynCalc, we square the amplitude and sum and average over spins. Finally, we numerically integrate over the three-body phase space to determine the partial width. This value is then multiplied by the mean lifetime of the $\tau$ to determine the branching ratio, which is then compared to the experimental upper limit $\text{Br}(\tau \rightarrow \mu\mu\mu) < 2.1 \times 10^{-8}$ \cite{Patrignani:2016xqp}.  

Potential constraints from $\tau^+ \rightarrow \mu^+ e^+ e^-$ are determined in a completely analogous fashion. Although, for this decay, only the photonic and $Z$ penguin diagrams contribute, whose amplitudes are similar to equations \ref{eq:taumumumu_photon} and \ref{eq:taumumumu_Z}, respectively, with appropriate modifications. The branching ratio is again computed with the assistance of FeynCalc and the result is compared with the experimental upper
limit $\text{Br}(\tau \rightarrow \mu e^+ e^-) < 1.8 \times 10^{-8}$ \cite{Patrignani:2016xqp}. 

The last type of process we consider involving taus is the decay $\tau \rightarrow K \nu$. This decay, which occurs in the Standard Model through a $W$ boson, can also potentially occur via a tree level exchange of a right-handed sbottom with two $\lambda'$ interactions. However, because we consider the right-handed sbottom to be heavy, we find no meaningful constraints from this decay.

\subsection{$B$ mesons}\label{sec:B_constraints}

Strong constraints on the  parameters in our model can be derived from $B_s - \bar{B}_s$ mixing. Particularly, the $\lambda'$ interactions induce $B_s - \bar{B}_s$ mixing via one loop box diagrams with either two right-handed sbottoms or two sneutrinos in the loop. Additionally, $B_s - \bar{B}_s$ mixing can also be induced by a one loop box diagram with two left-handed up squarks and two winos in the loop. It is useful to define the effective Lagrangian for this process
\begin{align}
\mathcal{L}_{\text{eff}} = C_{B_s}^{\text{SM}(\text{NP})} (\bar{s}\gamma^\alpha P_L b)(\bar{s}\gamma_\alpha P_L b) + \text{h.c.}
\end{align}
where $C_{B_s}^{\text{SM}(\text{NP})}$ is generated by the Standard Model (new physics). Explicitly, these are given by
\begin{align}
C_{B_s}^{\text{SM}} = -\frac{g^4}{128\pi^2 m_W^2}(V_{tb}V_{ts}^*)^2  S_0(x_t)
\end{align}
with $x_t = m_t^2 / m_W^2$, $m_t = m_t(m_t) \approx 162.3 \ \text{GeV}$, and $S_0(x_t) = \frac{x_t(4-11x_t+x_t^2)}{4(1-x_t)^2} - \frac{3 x_t^3 \log(x_t)}{2(1-x_t)^3} \approx 2.30$, and 
\begin{align}
C_{B_s}^{\text{NP}} = &\frac{1}{8}\lambda'_{i33}\lambda'^*_{i23}\lambda'_{j33}\lambda'^*_{j23}\frac{1}{i} \biggl(D_2[m^2_{\tilde{b}_R},m^2_{\tilde{b}_R},0,0] + D_2[m_{\tilde{\nu}_i}^2,m_{\tilde{\nu}_j}^2,m_b^2,m_b^2] \biggr) \notag \\
+ &\frac{g^4}{8}V_{ib}V_{is}^*V_{jb}V_{js}^* \frac{1}{i} D_2[m_{\tilde{u}_{Li}}^2,m_{\tilde{u}_{Lj}}^2,m_{\tilde{W}}^2,m_{\tilde{W}}^2].
\end{align}
In the limit of degenerate left-handed up squarks (which removes the wino contribution due to a super GIM mechanism) and $m_{\tilde{\nu}}^2 \gg m_b^2$, $C_{B_s}^{\text{NP}}$ simplifies to 
\begin{align}
C^{\text{NP}}_{B_s} = -\frac{\lambda'_{i33}\lambda'^*_{i23}\lambda'_{j33}\lambda'^*_{j23}}{128 \pi^2}\biggl(\frac{1}{m_{\tilde{b}_R}^2} + \frac{\log(m_{\tilde{\nu}_i}^2 / m_{\tilde{\nu}_j}^2)}{m_{\tilde{\nu}_i}^2 - m_{\tilde{\nu}_j}^2}\biggr).
\end{align}
Notice the $\lambda'$ dependence of this equation. The choice of parameters $\lambda'_{223}\lambda'_{233} > 0$ and $\lambda'_{323}\lambda'_{333} < 0$, initially motivated to achieve large values for $C_{LL}^\mu$, causes cancellations amongst the various diagrams. This is an example of the cancellations mentioned at the very end of section \ref{sec:calculations}. Importantly, these cancellations help lessen the constraints coming from $B_s - \bar{B}_s$ mixing. Again, we consider this a feature of the model, as the choice of parameters which lead to these cancellations is what is precisely preferred by $C_{LL}^\mu$. To constrain the relevant parameters, we follow the UT$fit$ collaboration \cite{Bona:2007vi} and define
\begin{align}
C_{B_s} e^{2i\phi_{B_s}} = \frac{\braket{B_s^0|H_{\text{eff}}^\text{full} |\bar{B}_s^0}}{\braket{B_s^0|H_{\text{eff}}^\text{SM} |\bar{B}_s^0}}.
\end{align}
We then have that $C_{B_s}$ and $\phi_{B_s}$ are given by
\begin{align}
C_{B_s} = \biggl|1 + \frac{C^{\text{NP}}_{B_s}}{C^{\text{SM}}_{B_s}} \biggr| \quad \text{and} \quad \phi_{B_s} = \frac{1}{2} \text{Arg}\biggl(1 + \frac{C^{\text{NP}}_{B_s}}{C^{\text{SM}}_{B_s}} \biggr).
\end{align}
The $2\sigma$ bounds on these two values, which can be found on the UT$fit$ collaboration's website, 
%\href{http://www.utfit.org/UTfit/ResultsSummer2016NP}{http://www.utfit.org/UTfit/ResultsSummer2016NP}, 
are given by $0.899 < C_{B_s} < 1.252$ and $-1.849^\circ < \phi_{B_s} < 1.959^\circ$. We find that, even with the cancellations between the diagrams, the constraint on $C_{B_s}$ still requires us to take $m_{\tilde{b}_R}$ and $m_{\tilde{\nu}}$ of order $10 \ \text{TeV}$ if we want the product $\lambda'_{223} \lambda'_{233}$ to be large.

The next decay we consider is $B \rightarrow K^{(*)} \nu \bar{\nu}$ which results from $b \rightarrow s \nu \bar{\nu}$. The quark level decay can potentially occur by a tree level exchange of a right-handed sbottom with two $\lambda'$ interactions. It is useful to define the effective Lagrangian for this process
\begin{align}
\mathcal{L}_{\text{eff}} = C_{b\rightarrow s \nu_i \bar{\nu}_j}^{\text{SM}(\text{NP})} (\bar{s}\gamma^\alpha P_L b)(\bar{\nu}_i\gamma_\alpha P_L \nu_j) + \text{h.c.}
\end{align}
where $C_{b\rightarrow s \nu_i \bar{\nu}_j}^{\text{SM}(\text{NP})}$ is generated by the Standard Model (new physics). Explicitly, these are given by
\begin{align}
C_{b\rightarrow s \nu_i \bar{\nu}_j}^{\text{SM}} = -\delta_{ij} \frac{g^4}{16\pi^2 m_W^2}V_{tb}V_{ts}^* X_0(x_t)
\end{align}
with $x_t$ defined as before and $X_0(x_t) = \frac{x_t(x_t+2)}{8(x_t-1)} + \frac{3x_t(x_t-2)}{8(x_t-1)^2}\log(x_t) \approx 1.48$, and
\begin{align}
C_{b\rightarrow s \nu_i \bar{\nu}_j}^{\text{NP}} = \frac{\lambda'_{j33}\lambda'^*_{i23}}{2 m^2_{\tilde{b}_R}}.
\end{align}
Next, consider the ratio $R_{B \rightarrow K^{(*)} \nu \bar{\nu}} = \Gamma^{\text{SM+NP}}(B \rightarrow K^{(*)} \nu \bar{\nu})/\Gamma^{\text{SM}}(B \rightarrow K^{(*)} \nu \bar{\nu})$. In terms of $C_{b\rightarrow s \nu_i \bar{\nu}_j}^{\text{SM}}$ and $C_{b\rightarrow s \nu_i \bar{\nu}_j}^{\text{NP}}$, it is given by
\begin{align}
R_{B \rightarrow K^{(*)} \nu \bar{\nu}} = \frac{\sum\limits_{i=1}^3\bigl|C_{b\rightarrow s \nu_i \bar{\nu}_i}^{\text{SM}} + C_{b\rightarrow s \nu_i \bar{\nu}_i}^{\text{NP}} \bigr|^2 + \sum\limits_{i,j=1}^3(1-\delta_{ij})\bigl|C_{b\rightarrow s \nu_i \bar{\nu}_j}^{\text{NP}} \bigr|^2}{\sum\limits_{i=1}^3\bigl|C_{b\rightarrow s \nu_i \bar{\nu}_i}^{\text{SM}}  \bigr|^2}.
\end{align}
The Belle search \cite{Grygier:2017tzo} provides 90\% CL upper bounds $R_{B \rightarrow K \nu \bar{\nu}} < 3.9$ and $R_{B \rightarrow K^* \nu \bar{\nu}} < 2.7$ on these ratios. We determine constraints on our parameter space from the limit on $R_{B \rightarrow K^* \nu \bar{\nu}}$.

Another potentially constraining process is the decay $B \rightarrow X_{\bar{s}} \gamma$ corresponding to the decay $\bar{b} \rightarrow \bar{s} \gamma$. The amplitude for the quark level process is given in equation \ref{eq:bsgamma} where, due to the photon being on-shell, only the dipole term contributes. We see that this amplitude depends on $A_{b2}^L$, given in equation \ref{eq:bsgammaA2L}, which is itself proportional to the inverse squared masses of the right-handed sbottom and sneutrinos. Because we take these particles to be heavy, we find no constraints from these decays. 

Finally, we also examined the decays $B \rightarrow \tau \nu$ and $B \rightarrow \mu \nu$. Both these decays occur in the Standard Model through a $W$ boson, although the latter decay is highly suppressed due to angular momentum conservation. They can also potentially occur as a result of a tree level right-handed sbottom exchange with two $\lambda'$ interactions. However, because we take the mass of the right-handed sbottom to be heavy, we find no constraints from these two decays.

\subsection{$Z$ decays}\label{sec:Zdecays}

\begin{figure}[t!]
  \centering
  \begin{subfigure}[b]{0.47\textwidth}
    \centering
    \includegraphics[width=\textwidth, bb = 0 0 401 163]{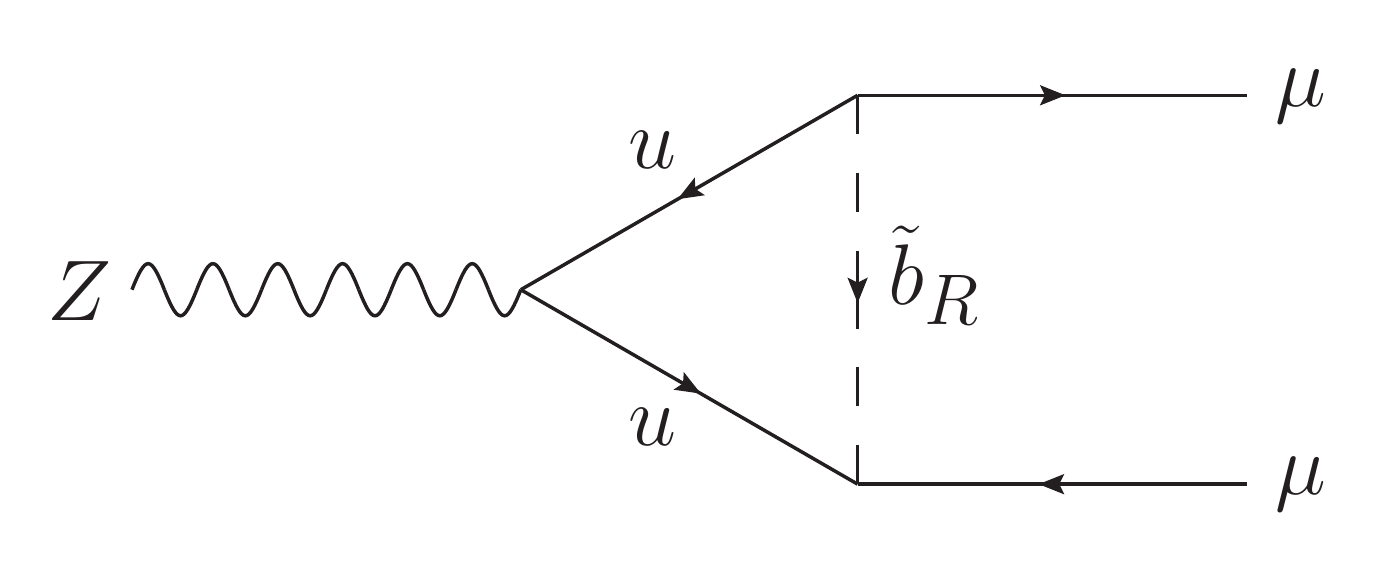}
    \caption{}
    \label{fig:Z_width_u}
  \end{subfigure}
  ~
  \begin{subfigure}[b]{0.47\textwidth}
    \centering
    \includegraphics[width=\textwidth, bb = 0 0 401 163]{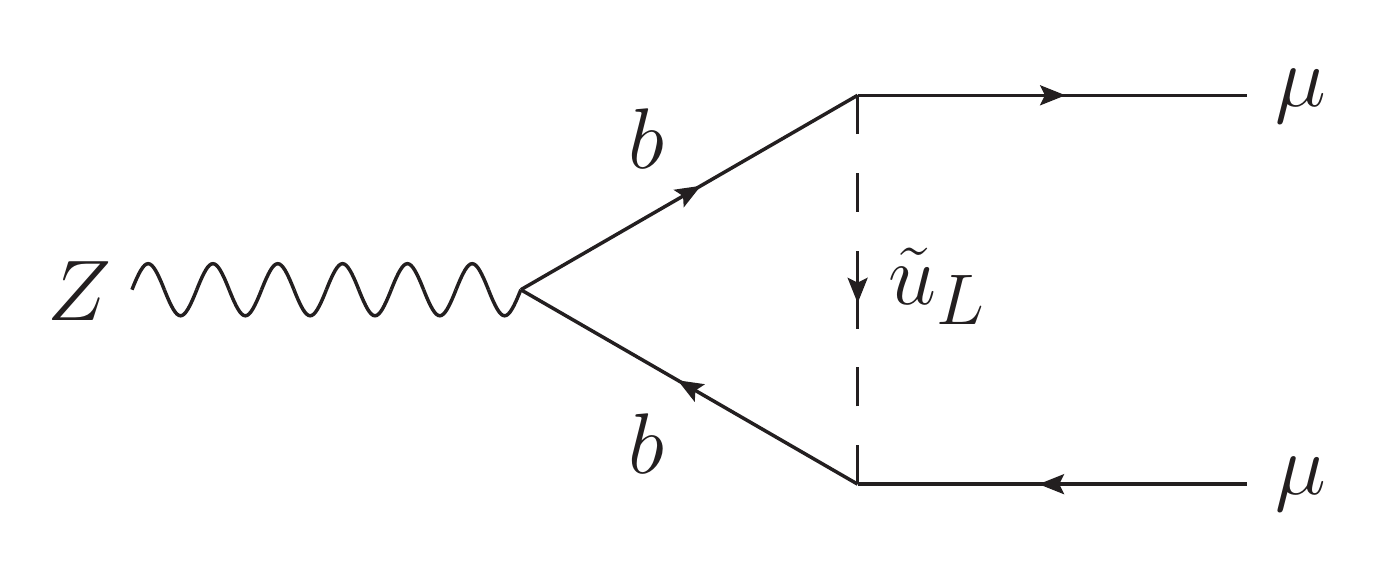}
    \caption{}
    \label{fig:Z_width_d}
  \end{subfigure}
  \caption{Example one loop Feynman diagrams contributing to $Z \rightarrow \mu \mu$.}\label{fig:Z_width}
\end{figure}

Loop level processes involving the right-handed sbottom and left-handed up squarks can potentially shift the partial width of the $Z$ to same flavour charged leptons or induce $Z$ decays to opposite flavour charged leptons. Example one loop Feynman diagrams are shown in figure \ref{fig:Z_width}. These diagrams contribute to the amplitude
\begin{align}
i \mathcal{M} = i \frac{g}{32\pi^2 c_W} B_{ij} \epsilon^\alpha \bar{u}_{e_i}\gamma_\alpha P_L v_{e_j} \label{eq:Zamp}
\end{align}
where $B_{ij} = B^1_{ij} + B^2_{ij} + B^3_{ij}$ and
\begin{align}
B^1_{ij} &= \sum_{l=1}^2 \tilde{\lambda}'_{jl3} \tilde{\lambda}'^*_{il3} \frac{m_Z^2}{m^2_{\tilde{b}_R}} \biggl[\biggl(1 - \frac{4}{3}s_W^2 \biggr)\biggl(\log\biggl(\frac{m_Z^2}{m^2_{\tilde{b}_R}}\biggr) - i \pi - \frac{1}{3} \biggr) + \frac{s_W^2}{9} \biggr], \label{eq:Zamp1} \\
B^2_{ij} &= 3 \tilde{\lambda}'_{j33} \tilde{\lambda}'^*_{i33} \biggl\{\frac{m_t^2}{m^2_{\tilde{b}_R}}\biggl(-\log\biggl(\frac{m_t^2}{m^2_{\tilde{b}_R}} \biggr) -1 \biggr) \notag \\
& + \frac{m_Z^2}{18 m^2_{\tilde{b}_R}}\biggl[(11 - 10 s_W^2) + (6 - 8 s_W^2)\log\biggl(\frac{m_t^2}{m^2_{\tilde{b}_R}} \biggr) + \frac{1}{10}(-9 + 16 s_W^2)\frac{m_Z^2}{m_t^2} \biggr] \biggl\}, \label{eq:Zamp2} \\
B^3_{ij} &= \sum_{l=1}^3 \tilde{\lambda}'_{jl3} \tilde{\lambda}'^*_{il3} \frac{m_Z^2}{m^2_{\tilde{u}_{Ll}}} \biggl[ \biggl(-\frac{2}{3}s_W^2 \biggr)\biggl(\log\biggl(\frac{m_Z^2}{m_{\tilde{u}_{Ll}}^2}\biggr) -i\pi - \frac{1}{2}\biggr) + \biggl(-\frac{1}{6} +\frac{1}{9}s_W^2 \biggr)\biggr]. \label{eq:Zamp3}
\end{align}
The function $B^1_{ij}$ is the contribution from the diagrams with a right-handed bottom squark and an up or charm quark in the loop. The function $B^2_{ij}$ is the contribution from the diagrams with a right-handed bottom squark and a top quark in the loop. These two functions match the results presented in \cite{Bauer:2015knc}, although we have retained additional terms in $B^2_{ij}$. The final function $B^3_{ij}$ is the contribution from the diagrams with a left-handed up squark and a bottom quark in the loop.

For the decays $Z \rightarrow \mu \mu$ and $Z \rightarrow \tau \tau$, we derive bounds by demanding that the interference term in the partial width computation between the Standard Model tree level diagram and the one loop contribution presented above is less than twice the experimental uncertainty on the partial width as given in \cite{Patrignani:2016xqp}. This leads to the bounds
\begin{align}
|\text{Re}[B_{22}]| < 0.32 \quad \text{and} \quad |\text{Re}[B_{33}]| < 0.39.
\end{align}
The decays $Z \rightarrow \mu \tau$ are bounded by demanding that the one loop contribution does not lead to a branching ratio larger than the experimental upper limit $\text{Br}(Z \rightarrow \mu\tau) < 1.2 \times 10^{-5}$ \cite{Patrignani:2016xqp}. This results in the bound
\begin{align}
\sqrt{|B_{23}|^2 + |B_{32}|^2} < 2.1.
\end{align}

\subsection{Other possible decays}

The right-handed sbottom and $\lambda'$ couplings can also induce several different tree level decays of $D$ mesons. For example, potential constraints can be derived from examining the decay $D^0 \rightarrow \mu \mu$, the ratio of branching ratios $R^{(*)}_{D^+} = \text{Br}(D^+ \rightarrow \mu^+ \nu \bar{K}^{0(*)})/\text{Br}(D^+ \rightarrow e^+ \nu \bar{K}^{0(*)})$ and $R_{D^0} = \text{Br}(D^0 \rightarrow \mu^+ \nu \bar{K}^-)/\text{Br}(D^0 \rightarrow e^+ \nu \bar{K}^-)$, and the decays $D_s \rightarrow \tau \nu$ and $D_s \rightarrow \mu \nu$. However, because we take the mass of the right-handed sbottom to be large, we find that none of these processes constrain our parameter space. 

The last type of processes we consider are upsilon decays to charged lepton pairs, $\Upsilon(1S) \rightarrow  e_i^- e_j^+$. The corresponding quark level process $b\bar{b} \rightarrow e_i^- e_j^+$ can potentially be induced by a tree level exchange of left-handed up squarks and two $\lambda'$ interactions. Integrating out the left-handed up squarks, we are left with the following effective Lagrangian 
\begin{align}
\mathcal{L}_{\text{eff}} = -\frac{\tilde{\lambda}'_{jl3}\tilde{\lambda}'^*_{il3}}{2 m_{\tilde{u}_{Ll}}^2} (\bar{b} \gamma^\alpha P_R b)(\bar{e}_i \gamma_\alpha P_L e_j).
\end{align}
Using this effective Lagrangian we can compute the branching ratio for the decay $\Upsilon \rightarrow \mu \tau$ as well potential modifications to the ratio of branching ratios $\text{Br}(\Upsilon \rightarrow \mu \mu)/\text{Br}(\Upsilon \rightarrow e e)$ and $\text{Br}(\Upsilon \rightarrow \tau \tau)/\text{Br}(\Upsilon \rightarrow e e)$. However, we find that the experimental upper limit on $\text{Br}(\Upsilon \rightarrow \mu \tau)$ is not stringent enough and that the decays $\Upsilon \rightarrow e_i^- e_i^+$ are not measured precisely enough to give any constraints on our parameter space.

\subsection{Collider searches}

The next type of constraint we discuss is direct LHC searches for pair produced up squarks subsequently decaying by $\lambda'$ interactions. Provided the up squarks are light enough, this process at the LHC would look like $pp \rightarrow \tilde{u}_{L}\tilde{u}^*_{L} \rightarrow \ell^+ \ell^- j j$ where, in our case, the two individual leptons can be either muons or taus, and both jets are b-jets. Thus, the possible signatures are two opposite sign muons, an opposite sign muon and tau pair, or two opposite sign taus, together with two b-jets.

There have been several ATLAS and CMS searches looking for these types of topologies, of which one of the most recent is \cite{Aaboud:2017opj}. This is an ATLAS search with centre of mass energy $\sqrt{s} = 13 \ \text{TeV}$ and integrated luminosity $36.1 \ \text{fb}^{-1}$. It considers stop pair production with the stops decaying by $\lambda'$ interactions. The final state topologies it considers are $\ell^+ \ell^- j j$ where $\ell = e$ or $\mu$ and both jets are b-jets. The search presents lower limits for stop masses in the $\text{Br}(\tilde{t}\rightarrow be) + \text{Br}(\tilde{t}\rightarrow b\mu) + \text{Br}(\tilde{t}\rightarrow b\tau) = 1$ plane. To extract limits from this search, we first make the simplifying assumption that the efficiencies to pass the cuts (which require one of $ee$, $e\mu$, or $\mu\mu$) are zero if either stop decays to a $\tau$ and a $b$. Then, using the exclusion plot, the provided $95\%$ CL upper limit on the number of BSM signals, and the stop pair production cross section which we compute using NNLL-fast \cite{Beenakker:2016lwe, Beenakker:1997ut, Beenakker:2010nq, Beenakker:2016gmf}, we can determine the efficiencies for both stops decaying to a $\mu$ and $b$. Once we have the efficiencies, determining limits on our model is straightforward. 

To do this, we first determine the production cross sections for the three individual up squarks. For simplicity, we use the pair production cross section for stops for the first two generations as well. This is equivalent to assuming a heavy gluino. Then, we compute the branching ratios for our up squarks to decay to a $\mu$ and a $b$. Here, we consider the decays $\tilde{u}_L \rightarrow \mu b$, $\tilde{u}_L \rightarrow \tau b$, and $\tilde{u}_L \rightarrow \tilde{W} q$ where the last decay includes both neutral and charged winos. For large values of the $\lambda'$ couplings $(\lambda' \gtrsim 1)$, the first two decays dominate. We then compute the number of expected signals by multiplying the integrated luminosity, the cross sections, the efficiencies, and the squared branching ratios for $\tilde{u}_L \rightarrow \mu b$. Comparing this number to the provided $95\%$ CL upper limit on the number of BSM signals, we determine whether points in parameter space are excluded.

It is worth mentioning that we also examined experimental searches looking for the final state $\tau^+ \tau^- j j$ where again the jets are b-jets. One of the most recent searches looking for this final state is the CMS search \cite{Sirunyan:2017yrk}. However, this search fails to provide any additional constraints in the parameter space we examine. This is simply because these types of searches provide weaker limits than searches looking for $\ell^+ \ell^- j j$ ($\ell = e$ or $\mu$) due to the difficulty in reconstructing taus. 

\subsection{Landau poles}

To generate large values of $C_{LL}^\mu$, we will need to take the four $\lambda'$ couplings under consideration to be fairly large. This will then result in Landau poles below the Planck scale. To calculate the energy scales of these Landau poles, we use the following procedure. First, we evolve the three gauge couplings and the top, bottom, and tau Yukawa couplings up to the left-handed up squarks mass scale using the Standard Model beta functions. From there, we evolve these parameters and the four $\lambda'$ couplings up to the right-handed sbottom and sneutrino mass scale. The beta functions used for this evolution are the one loop RPVMSSM beta functions \cite{Allanach:1999mh}, except with the following modification. As some of the sparticle masses are at the very top of this evolution scale, we remove their effects on the beta functions. Precisely, we remove the effects on the beta functions due to the sfermions coming from the superfields $U^c$, $D^c$, $L$, and $E^c$. Lastly, we evolve the parameters upwards from this scale using the full one loop RPVMSSM beta functions and determine the Landau pole accordingly.

\section{Results}\label{sec:results}

\begin{figure}[t!]
  \centering
  \begin{subfigure}[b]{0.47\textwidth}
    \centering
    \includegraphics[width=\textwidth, bb = 0 0 729 725 ]{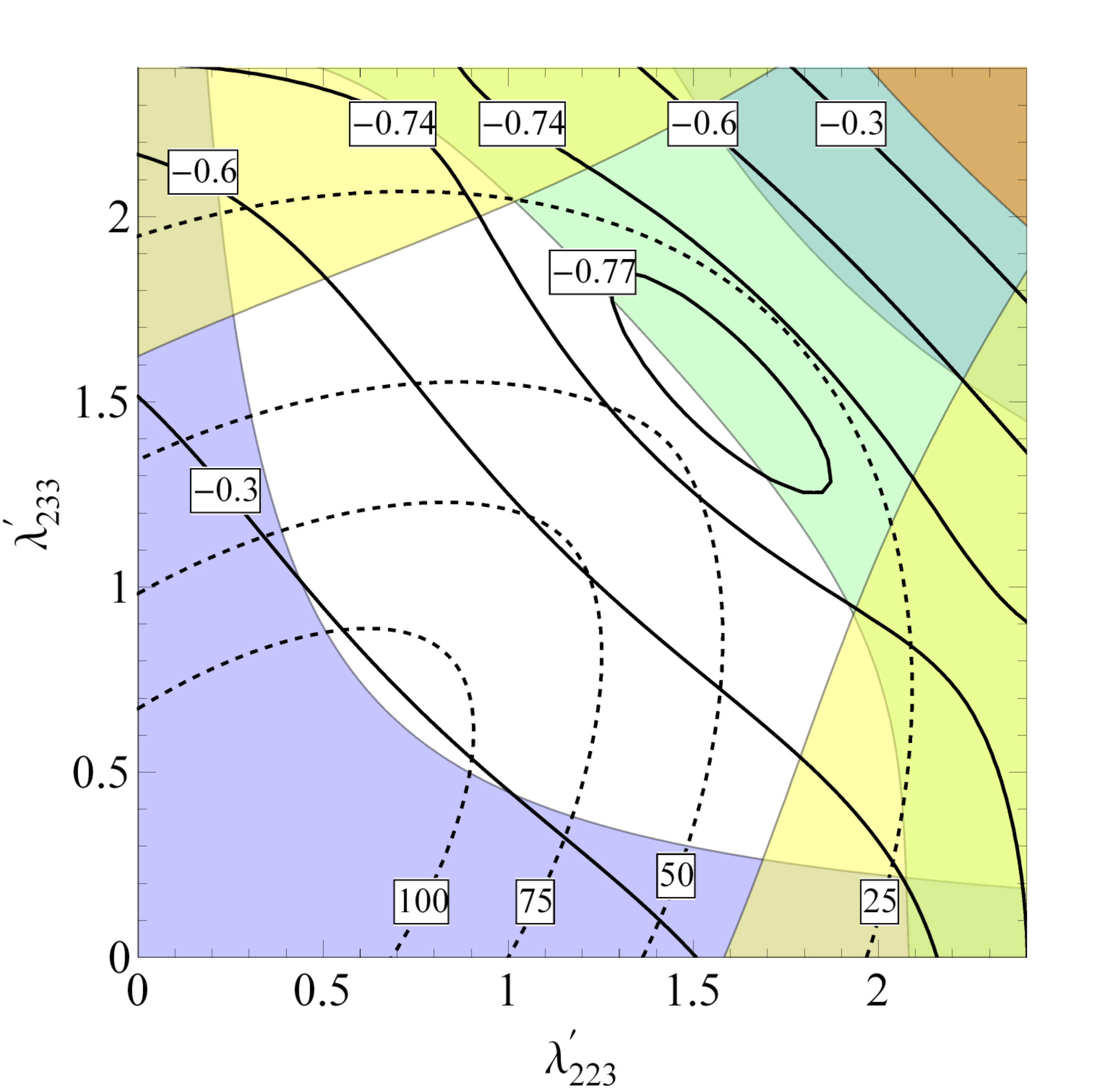}
    \caption{}
    \label{fig:223_233}
  \end{subfigure}
  ~
  \begin{subfigure}[b]{0.47\textwidth}
    \centering
    \includegraphics[width=\textwidth, bb = 0 0 729 725 ]{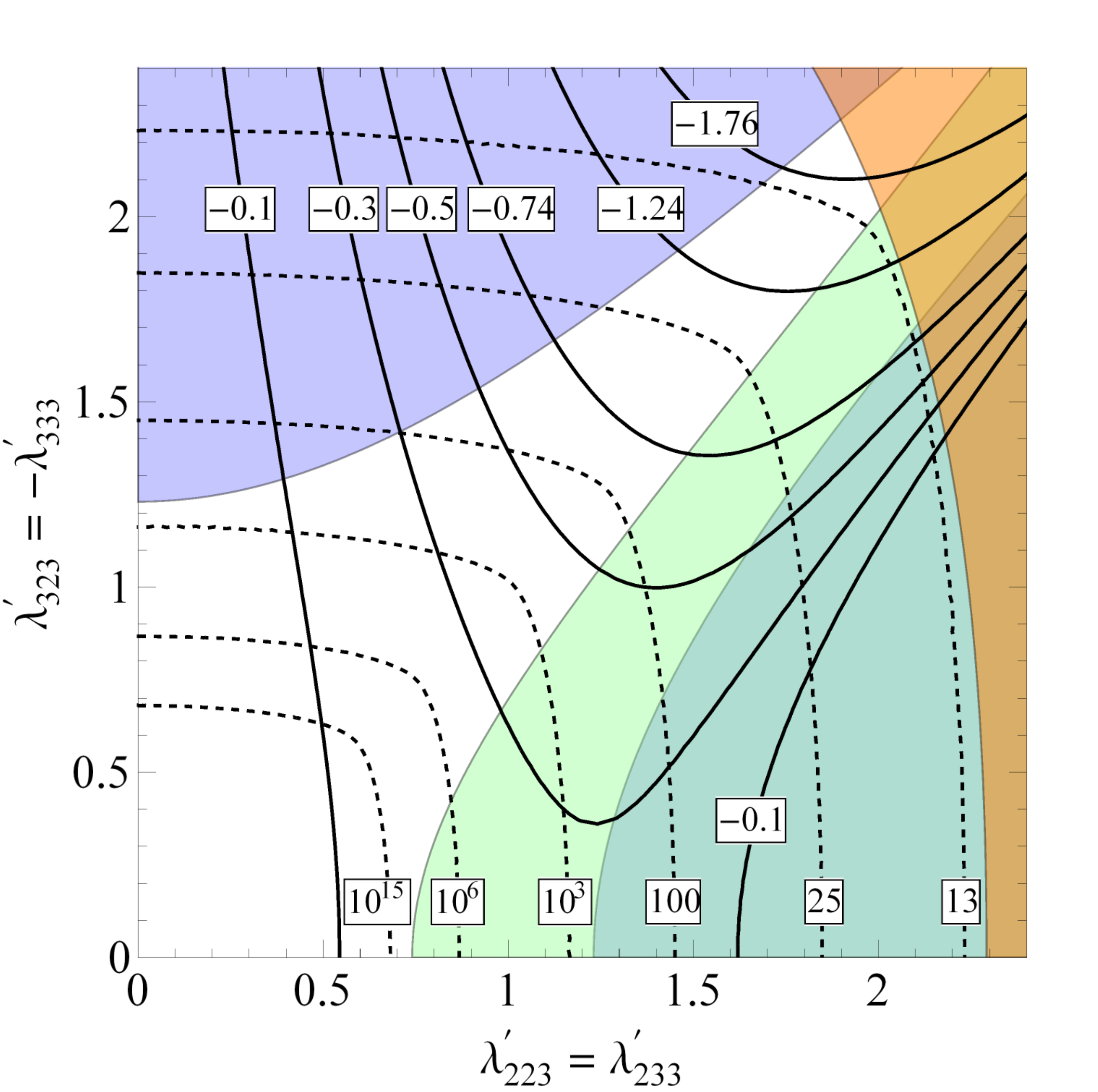}
    \caption{}
    \label{fig:223233_323333}
  \end{subfigure}
  ~
  \begin{subfigure}[b]{0.47\textwidth}
    \centering
    \includegraphics[width=\textwidth, bb = 0 0 729 726 ]{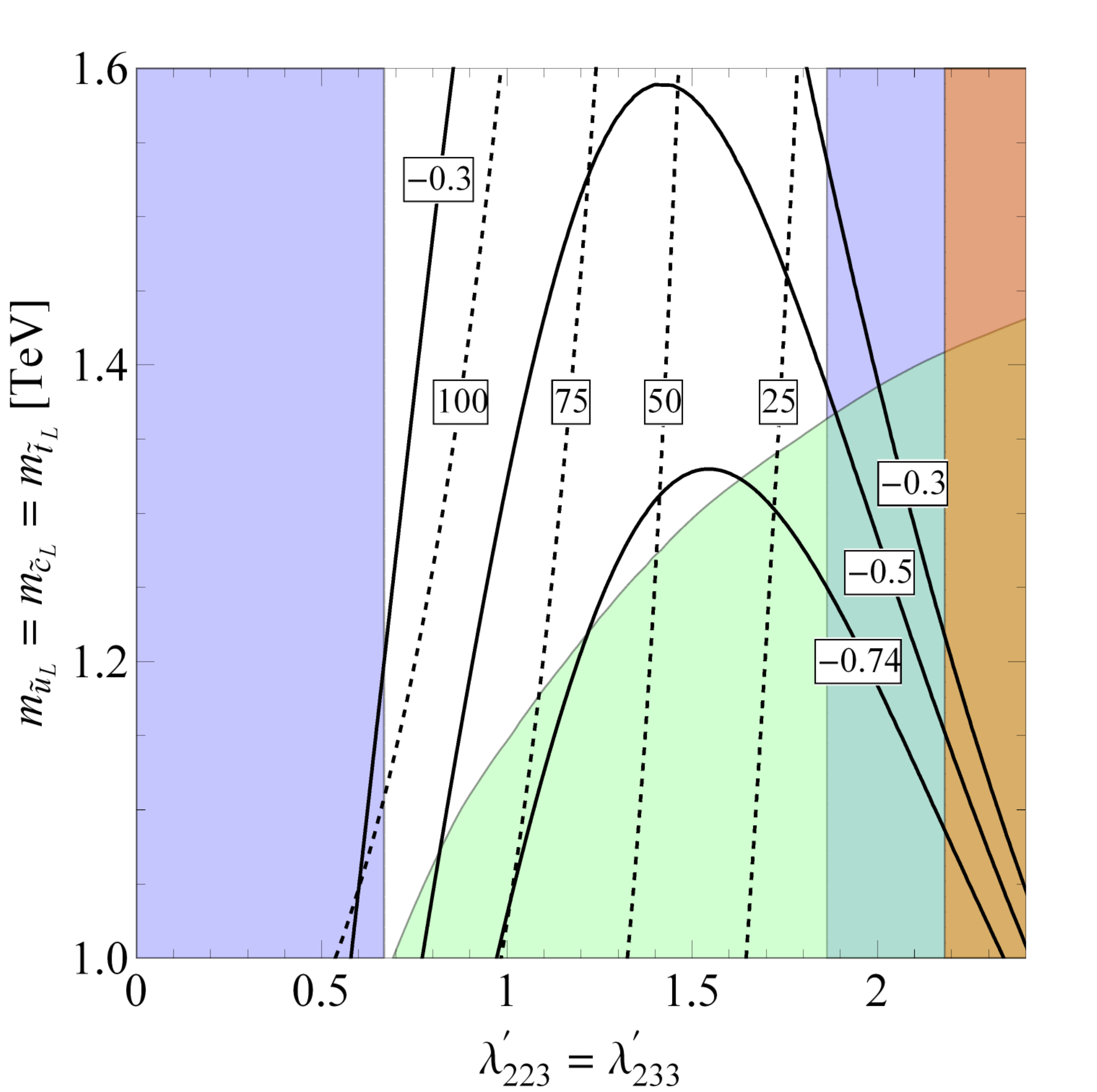}
    \caption{}
    \label{fig:223233_mus}
  \end{subfigure}
   ~
  \begin{subfigure}[b]{0.47\textwidth}
    \centering
    \includegraphics[width=\textwidth, bb = 0 0 729 731 ]{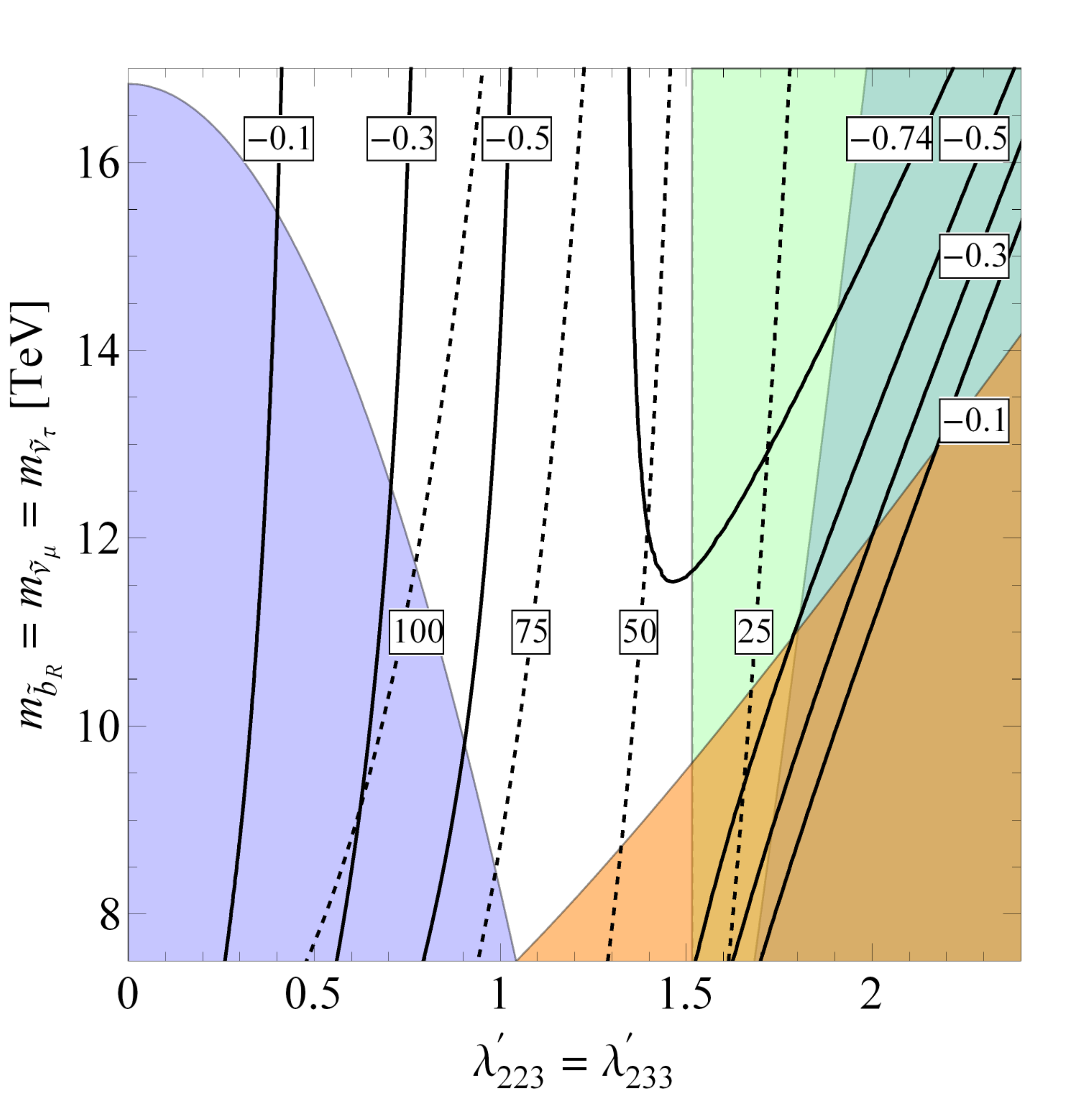}
    \caption{}
    \label{fig:223233_mds}
  \end{subfigure}
  \caption{Four example figures showing solid contours of $C_{LL}^\mu$. For figure \ref{fig:223_233}, we set $\lambda'_{323} = -\lambda'_{333} = 1.4$, $m_{\tilde{W}} = 300 \ \text{GeV}$, $m_{\tilde{u}_L} = m_{\tilde{c}_L} = m_{\tilde{t}_L} = 1.3 \ \text{TeV}$, and $m_{\tilde{b}_R} = m_{\tilde{\nu}_\mu} = m_{\tilde{\nu}_\tau} = 13 \ \text{TeV}$. For figure \ref{fig:223233_323333}, the masses are set to the same values as in figure \ref{fig:223_233}. For figures \ref{fig:223233_mus} and \ref{fig:223233_mds}, $\lambda'_{323}$, $\lambda'_{333}$, and the masses not being varied are again set to the values used in figure \ref{fig:223_233}. Dashed contours show energy scales of Landau poles in TeV. Parameter space excluded by $\tau \rightarrow \mu \mu \mu$ is shown in yellow. Parameter space excluded by $B_s - \bar{B}_s$ mixing is shown in blue. Parameter space excluded by $B \rightarrow K^{(*)} \nu \bar{\nu}$ is shown in orange. Finally, parameter space excluded by direct LHC searches is shown in green.}\label{fig:results}
\end{figure}

Our results are presented in the four plots in figure \ref{fig:results}. In these plots, we show solid contours of constant values of $C_{LL}^\mu$. Also shown are dashed contours representing energy scales in $\text{TeV}$ at which Landau poles occur. In addition, we find that relevant parameter space is excluded by the processes $\tau \rightarrow \mu\mu\mu$, $B_s - \bar{B}_s$ mixing, and $B \rightarrow K^{(*)} \nu \bar{\nu}$, as well as direct LHC searches. In making these plots, we have taken $\lambda'_{223}$, $\lambda'_{233}$, and $\lambda'_{323}$ positive and $\lambda'_{333}$ negative. There are essentially identical plots with $\lambda'_{223} < 0$ and $\lambda'_{233} < 0$ or $\lambda'_{323} < 0$ and $\lambda'_{333} > 0$. For each of these plots, we have set the mass of the wino to be $300 \ \text{GeV}$. We have also only considered mass degenerate left-handed up squarks and we have set the mass of the right-handed sbottom equal to the masses of the sneutrinos. Since we are primarily interested in the wino diagrams contribution to $C_{LL}^\mu$ we vary the parameters $\lambda'_{223}$ and $\lambda'_{233}$ in each of the plots. These are the only parameters varied in figure \ref{fig:223_233}, while we also vary $\lambda'_{323}$ and $\lambda'_{333}$ in figure \ref{fig:223233_323333}, the masses of the left-handed up squarks in figure \ref{fig:223233_mus}, and masses of the right-handed sbottom and the sneutrinos in figure \ref{fig:223233_mds}.

Examining the plots, we observe the following features. First, it is difficult to generate very large values of $C_{LL}^\mu$ in this setup. We see that in all four plots only a small portion of the unexcluded parameter space has $C_{LL}^\mu < -0.74$, the upper limit of the $2\sigma$ region capable of explaining the anomalies as stated in Ref.\ \cite{Capdevila:2017bsm}. Indeed, in figures \ref{fig:223_233}, \ref{fig:223233_mus}, and \ref{fig:223233_mds} the largest value of $C_{LL}^\mu$ which can be generated and is not excluded is $\approx -0.77$. We see in figure \ref{fig:223233_323333} that larger values of $C_{LL}^\mu$ can be generated but only if all four $\lambda'$ couplings are taken large in magnitude. This leads to the second feature, large values of $C_{LL}^\mu$ necessarily imply low scale Landau poles. For each plot, the parameter region with $C_{LL}^\mu < -0.74$ also has a Landau pole at an energy scale $\lesssim 70 \ \text{TeV}$. In fact, in figure \ref{fig:223233_323333} we see that a portion of the otherwise unexcluded parameter space has Landau poles at energy scales less than the masses of the right-handed sbottom and sneutrinos. This region is thus also excluded. Third, notice that in figure \ref{fig:223233_323333} the two regions excluded by $B_s - \bar{B}_s$ mixing do not converge, even for the largest values of the $\lambda'$ couplings. This is an example of the cancellation amongst diagrams discussed in section \ref{sec:B_constraints}. Fourth, as shown by figure \ref{fig:223233_mus}, the direct LHC search constraints require the masses of the left-handed up squarks to be $\gtrsim 1.4 \ \text{TeV}$ if these particles decay only to $\mu b$. Smaller masses are allowed provided the left-handed up squarks decay to $\tau b$ as well. The final feature we wish to mention is that, as shown in figure \ref{fig:223233_mds}, the masses of the right-handed sbottom and sneutrinos need to be $\gtrsim 7.5 \ \text{TeV}$. This demonstrates the smallest mass splitting between these particles and the left-handed up squarks that we can achieve in this setup.

\subsection{Additional remarks}

\begin{figure}[t!]
  \centering
  \includegraphics[width=0.47\textwidth, bb = 0 0 729 729 ]{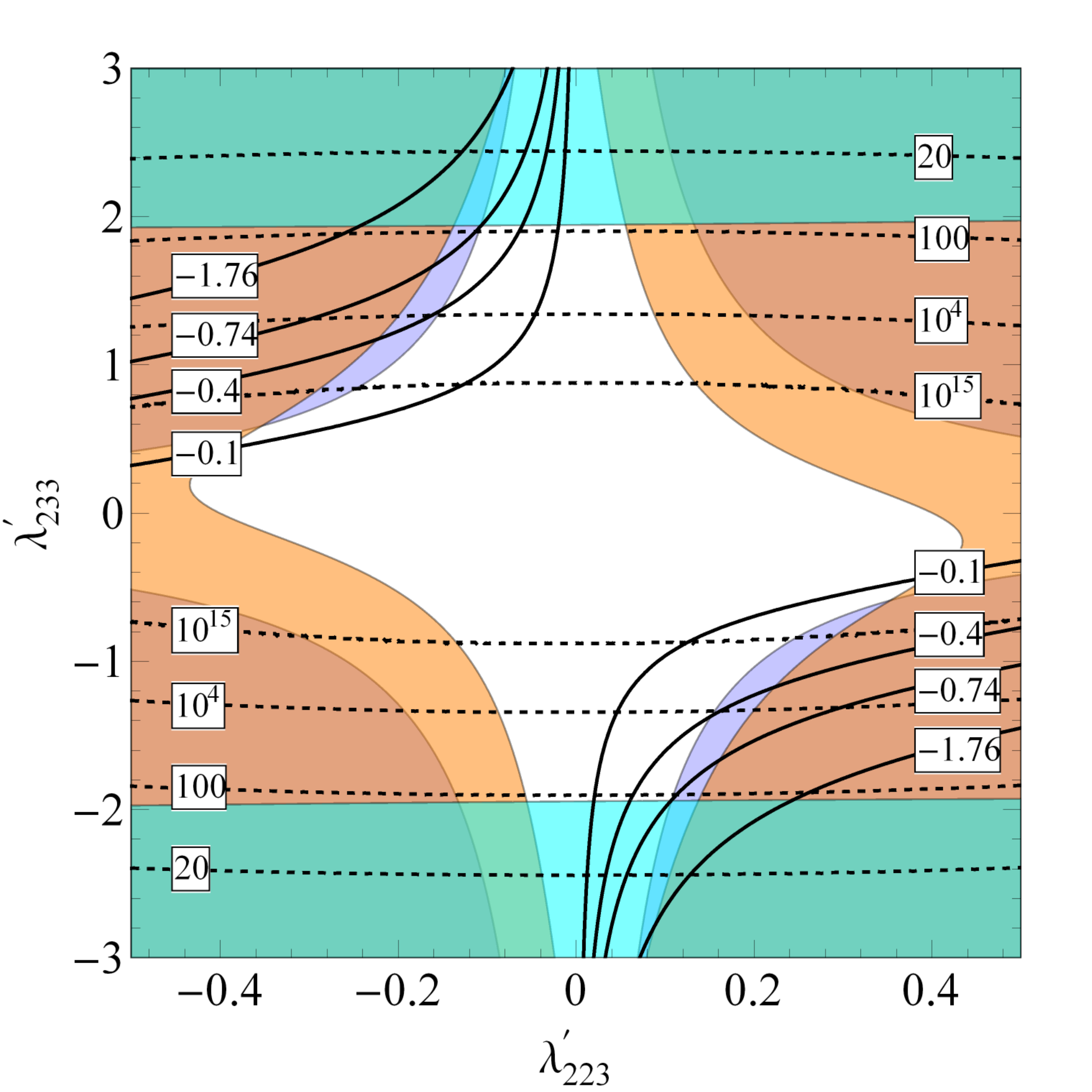}
  \caption{Example figure showing solid contours of $C_{LL}^\mu$ for parameter space similar to that considered in Ref.\ \cite{Das:2017kfo}. For this figure, we set $\lambda'_{323} = 0.05$, $\lambda'_{333} = -0.5$, $m_{\tilde{W}} = 300 \ \text{GeV}$, and $m_{\tilde{u}_L} = m_{\tilde{c}_L} = m_{\tilde{t}_L} = m_{\tilde{b}_R} = m_{\tilde{\nu}_\mu} = m_{\tilde{\nu}_\tau} = 2 \ \text{TeV}$. Dashed contours show energy scales of Landau poles in TeV. Parameter space excluded by $B_s - \bar{B}_s$ mixing is shown in blue. Parameter space excluded by $B \rightarrow K^{(*)} \nu \bar{\nu}$ is shown in orange. Finally, parameter space excluded by $Z$ decays to charged leptons is shown in cyan. }
  \label{fig:223_233_ps2}
\end{figure}

It is interesting to compare our results with those in Ref.\ \cite{Das:2017kfo}. There, the masses of all the sparticles are at the $\text{TeV}$ scale and the negative contributions to $C_{LL}^\mu$ come from the four-$\lambda'$ loop diagrams. In figure \ref{fig:223_233_ps2}, we show an example plot examining this parameter space. In this figure, we have set $\lambda'_{323} = 0.05$, $\lambda'_{333} = -0.5$, $m_{\tilde{W}} = 300 \ \text{GeV}$, and $m_{\tilde{u}_L} = m_{\tilde{c}_L} = m_{\tilde{t}_L} = m_{\tilde{b}_R} = m_{\tilde{\nu}_\mu} = m_{\tilde{\nu}_\tau} = 2 \ \text{TeV}$. By setting the masses of the left-handed up squarks and right-handed bottom squark to $2 \ \text{TeV}$ we avoid potential constraints from direct LHC searches.\footnote{As shown in figure \ref{fig:223233_mus}, the limits from pair produced squarks decaying to $\mu\mu b b$ saturate at $\sim 1.4 \ \text{TeV}$. Additionally, pair produced right-handed sbottoms can also decay to the final state $\nu \nu b b$. The limits from this type of signature saturate at $\sim 1.1 \ \text{TeV}$ \cite{CMS-PAS-SUS-18-001}.} A new feature in this figure compared to the plots in figure \ref{fig:results} is that some of the parameter space is excluded by $Z$ decays to charge leptons. This type of constraint was not considered in Ref.\ \cite{Das:2017kfo}. Further, we see that achieving values of $C_{LL}^\mu < -0.74$ is still difficult in this setup as well. Also, the energy scales of the Landau poles are similar to those in figure \ref{fig:results}. Finally, we note that by setting the masses of the sparticles to be of the same order, we are required to consider a hierarchical structure for the four $\lambda'$ couplings under consideration. In our setup, the $\lambda'$ couplings can be of the same magnitude but we are forced to consider a hierarchical structure for the sparticle masses.

To generate large values of $C_{LL}^\mu$ we have considered large values for the four parameters $\lambda'_{223}$, $\lambda'_{233}$, $\lambda'_{323}$, and $\lambda'_{333}$. Moreover, these couplings should also generate contributions to $C_{LL}^{\tau}$, $C_{LL}^{\tau \mu}$, and $C_{LL}^{\mu \tau}$, each defined analogously to $C_{LL}^\mu$
\begin{align}
\mathcal{H}_{\text{eff}} = - \frac{4G_F}{\sqrt{2}}V_{tb}V_{ts}^* \frac{\alpha}{4\pi} C_{LL}^{ij} (\bar{s}\gamma_\alpha P_L b)(\bar{\ell}_{i}\gamma^\alpha P_L \ell_j) + \text{h.c.}
\end{align}
with $C_{LL}^{i} \equiv C_{LL}^{ii}$. These operators result in decays such as $B_s \rightarrow \mu \tau$, $B_s \rightarrow \tau \tau$, $B \rightarrow K^{(*)} \mu \tau$, and $B \rightarrow K^{(*)} \tau \tau$. Generically, each of these decays are not measured precisely enough (or at all) to cause any potential conflicts. For example, using just the effective Hamiltonian above,\footnote{This is not quite right for the decay $B_s \rightarrow \tau \tau$ since there is also a Standard Model contribution given by $\text{Br}(B_s \rightarrow \tau \tau) = (7.73 \pm 0.49) \times 10^{-7}$ \cite{Bobeth:2013uxa, Bobeth:2014tza}.} we find
\begin{align}
\text{Br}(B_s \rightarrow \mu \tau) = 5.4 \times 10^{-9} (|C_{LL}^{\mu\tau}|^2 + |C_{LL}^{\tau\mu}|^2)
\end{align}
and
\begin{align}
\text{Br}(B_s \rightarrow \tau \tau) = 1.0 \times 10^{-8} |C_{LL}^{\tau}|^2.
\end{align} However, we are unaware of any experimental bound on the former decay\footnote{An indirect bound can be placed on $\text{Br}(B_s \rightarrow \mu \tau)$ by noting that this branching ratio is similar in size to $\text{Br}(B^+ \rightarrow K^+ \mu \tau)$ \cite{Becirevic:2016zri} and that the Babar search \cite{Lees:2012zz} has provided the bound $\text{Br}(B^+ \rightarrow K^+ \mu \tau) < 4.8 \times 10^{-5}$.} while the current experimental bound on the latter decay is $\text{Br}(B_s \rightarrow \tau \tau) < 6.8 \times 10^{-3}$ \cite{Aaij:2017xqt}. More details regarding the other two decays can be found in \cite{Crivellin:2015era, Capdevila:2017iqn}.

The $\lambda'$ couplings will also induce neutrino masses at the one loop level. Applying the general formula found in \cite{Barbier:2004ez} to our setup, we find contributions to the neutrino mass matrix given by
\begin{align}
M_{ij}^\nu = \frac{3}{16\pi^2} \lambda'_{i33} \lambda'_{jl3} m_b (\tilde{m}^{d \, 2}_{LR})_{l3} \frac{\log(m_{\tilde{b}_R}^2 / m_{\tilde{d}_{Ll}}^2)}{m_{\tilde{b}_R}^2 - m_{\tilde{d}_{Ll}}^2} + (i \leftrightarrow j) \label{eq:nu_masses}
\end{align}
where $\tilde{m}^{d \, 2}_{LR}$ is the left-right sdown mass mixing matrix. In the normal RPVMSSM, this will generate neutrino masses that are far too large. As an example, consider the contribution to the $i = j = 2$ entry from the case $l = 3$. We then have that $(\tilde{m}^{d \, 2}_{LR})_{33} = (A_b - \mu \tan\beta)m_b$. Taking $A_b - \mu \tan\beta = 1 \ \text{TeV}$, $m_{\tilde{b}_R} = 13 \ \text{TeV}$, $m_{\tilde{b}_L} = 1.3 \ \text{TeV}$, and $\lambda'_{233} = 1.2$, we find $M^\nu_{22} \sim 10 \ \text{keV}$, much larger than the $\sim 0.1 \ \text{eV}$ limit on the neutrino mass scale. This potential difficulty was also pointed out in Ref.\ \cite{Altmannshofer:2017poe}, who suggested $A_b$ and $\mu \tan \beta$ may cancel each other so that $\tilde{m}^{d \, 2}_{LR}$ is small. Another possibility mentioned in the same reference is that there may be additional unrelated contributions to the neutrino mass matrix which cancel those coming from equation \ref{eq:nu_masses}. Alternatively, the situation can be improved by assuming a model of supersymmetry that possesses a $U(1)_R$ symmetry identified with lepton number \cite{Frugiuele:2011mh,Frugiuele:2012pe,Gherghetta:2003he}. These types of models, which feature the $\lambda'$ couplings, assign different lepton number charges to the left and right-handed squarks. As a result, $\tilde{m}^{d \, 2}_{LR}$ vanishes in the limit that the $R$-symmetry is exact. However, the $R$-symmetry will be broken by at least anomaly mediation and this will generate contributions to $\tilde{m}^{d \, 2}_{LR}$ proportional to the gravitino mass. Parametrically we have
\begin{align}
(\tilde{m}^{d \, 2}_{LR})_{33} \sim  m_{3/2} \frac{m_b}{16\pi^2}
\end{align} 
and this leads to
\begin{align}
M^\nu_{22} \sim 0.1 \text{eV} \biggl(\frac{m_{3/2}}{1\text{GeV}} \biggr),
\end{align}
where we have used the same values for the parameters as before.  Thus, provided that the gravitino mass is lighter than $1 \ \text{GeV}$, the model is safe from bounds on neutrino masses. Note that a gravitino in that mass range and stable on cosmological time scales can be problematic for cosmology as it can overclose the universe \cite{Moreau:2001sr}. This can be solved by having a low reheat temperature or late entropy production.

Finally, we would like to briefly comment on the $R_{D^{(*)}}$ anomalies. These anomalies are the apparent enhancement of the ratio of branching ratios $R_D$ and $R_{D^*}$ defined in equation \ref{eq:R_D}. Specifically, the current experimental values for these ratios are \cite{Amhis:2016xyh}
\begin{align}
R_D = 0.403 \pm 0.040\text{(stat)} \pm 0.024\text{(syst)}  \ \text{and} \ R_{D^*} = 0.310 \pm 0.015\text{(stat)} \pm 0.008\text{(syst)}
\end{align}
while the Standard Model predicts \cite{Bernlochner:2017jka}
\begin{align}
R_D = 0.299 \pm 0.003 \quad \text{and} \quad R_{D^*} = 0.257 \pm 0.003.
\end{align} 
When combined, these measurements represent an approximate $4\sigma$ deviation away from the Standard Model \cite{Amhis:2016xyh}. The underlying quark transition $b \rightarrow c \ell \nu$ ($\ell = e$, $\mu$, or $\tau$) can potentially occur by a tree level exchange of a right-handed sbottom with two $\lambda'$ interactions. Indeed, the effect of these diagrams on the anomalies has previously been examined in the literature \cite{Deshpande:2012rr, Deshpand:2016cpw, Altmannshofer:2017poe}. Following the analysis in \cite{Deshpand:2016cpw}, we find that our setup has essentially no impact on these anomalies because we have taken the mass of the right-handed sbottom to be large. 

\section{Conclusion}\label{sec:conclusion}

In this paper, we examined the $b \rightarrow s \mu \mu$ anomalies within a supersymmetric framework with $R$-parity violation. Model independent analyses performed by different groups have shown that one way to explain these anomalies is to generate a negative contribution to the four-fermi operator $(\bar{s}\gamma_\alpha P_L b)(\bar{\mu}\gamma^\alpha P_L \mu)$. To do this, we considered the $R$-parity violating superpotential term $\lambda' LQD^c$ and studied many different diagrams. Initially, we examined a potentially relevant tree level diagram but found that it generates an effective four-fermi operator with an incorrect chirality structure. We then proceeded by studying multiple types of one loop diagrams. Specifically, we investigated the scenario in which the primary contribution is given by one loop box diagrams featuring a wino, with smaller contributions from one loop box diagrams featuring four $\lambda'$ interactions. This led us to turning on the couplings $\lambda'_{223}$, $\lambda'_{233}$, $\lambda'_{323}$, and $\lambda'_{333}$ with $\lambda'_{223}\lambda'_{233} > 0$ and $\lambda'_{323}\lambda'_{333} < 0$. Additionally, this scenario requires a spectrum in which the masses of the wino and left-handed up squarks are of order $1 \ \text{TeV}$ and the masses of the right-handed sbottom and sneutrinos are of order $10 \ \text{TeV}$. We then studied many physical processes relevant to our parameters. Constraints were derived from various $\tau$ decays including $\tau \rightarrow \mu$ meson, $\tau \rightarrow \mu \gamma$, $\tau \rightarrow \mu \mu \mu$, and $\tau \rightarrow \mu e^+ e^-$. Additional constraints were determined from $B_s - \bar{B}_s$ mixing, $B \rightarrow K^{(*)} \nu \bar{\nu}$, $Z$ decays to charged leptons, and direct LHC searches. Four example plots examining the parameter space were presented. These plots demonstrated that this setup can potentially explain the anomalies, although generating large contributions can be challenging. Moreover, to explain the anomalies, the four $\lambda'$ couplings each need to be large and this necessarily leads to low scale Landau poles. We then compared our setup with a more traditional supersymmetric spectrum in which the masses of all the sparticles are at the $\text{TeV}$ scale. Finally, we briefly discussed decays such as $B_s \rightarrow \mu \tau$ and $B_s \rightarrow \tau \tau$, contributions to the neutrino mass matrix, and how our model effects the anomalies related to the observables $R_{D^{(*)}}$.  

To conclude, we will briefly summarize the different potential solutions to the $b \rightarrow s \mu \mu$ anomalies and the $R_{D^{(*)}}$ anomalies which involve $R$-parity violation. We have found a new region of parameter space capable of potentially explaining the $b \rightarrow s \mu \mu$ anomalies. This region is characterized by the wino and left-handed up squarks having masses of order $1 \ \mathrm{TeV}$ and the right-handed sbottom and sneutrinos having masses of order $10 \ \text{TeV}$. The four couplings $\lambda'_{223}$, $\lambda'_{233}$, $\lambda'_{323}$, and $\lambda'_{333}$ are each of order $1$. For these parameters, $R_{D^{(*)}}$ receives no significant additional contributions, and thus this region of parameter space is unable to explain the anomalies associated with these observables. Crucially, this potential solution to the $b \rightarrow s \mu \mu$ anomalies relies on a light wino. If, on the other hand, the wino turns out to be heavy, then the $b \rightarrow s \mu \mu$ anomalies can still be explained as presented in \cite{Das:2017kfo}. This requires the masses of the left-handed up squarks, right-handed sbottom, and sneutrinos to be each of order $1 \ \mathrm{TeV}$. The same four $\lambda'$ as in the light wino case are again non-zero but now $\lambda'_{233}$ and $\lambda'_{333}$ are of order $1$ while $\lambda'_{223}$ and $\lambda'_{323}$ are much smaller. Although, as shown in figure \ref{fig:223_233_ps2}, totally explaining the $b \rightarrow s \mu \mu$ anomalies can still be challenging. These parameters can also lead to moderate contributions to $R_{D^{(*)}}$ \cite{Das:2017kfo}. Finally, it is possible to fully explain the anomalies in $R_{D^{(*)}}$ by making the mass of the right-handed sbottom less than $1 \ \mathrm{TeV}$ and only $\lambda'_{333}$ large, but in this case it is now difficult to also explain the $b \rightarrow s \mu \mu$ anomalies \cite{Deshpande:2012rr, Deshpand:2016cpw, Altmannshofer:2017poe}.

\acknowledgments
This work was supported in part by the Natural Sciences and Engineering Research Council of Canada (NSERC). KE acknowledges support from the Alexander Graham Bell Canada Graduate Scholarships Doctoral Program (CGS D) and from the Ontario Graduate Scholarship (OGS).

\bibliographystyle{JHEP}
\bibliography{Paper1}

\end{document}